\documentclass{emulateapj-rtx4}
\usepackage{apjfonts}
\usepackage{times}

\input psfig.tex

\newcommand{\kms}{km\,s$^{-1}$} 
\newcommand{\Msun}{M$_{\odot}$}
\newcommand{\Mcl}{{\cal{M}}_{\rm cl}}
\newcommand{\rh}{r_{\rm h}}
\newcommand{\Rgal}{R_{\rm gal}}
\newcommand{\cM}{{\cal{M}}}
\newcommand\colzero {\null}
\newcommand\cola {&}
\newcommand\colb {&}
\newcommand\colc {&}

\newcommand\cole {&}

\newcommand\colg {&}

\newcommand\coli {&}
\newcommand\eol{\\}

\newcommand{\picplace}[1]{\vbox{\hrule\@height 0.4pt\@width\hsize
\hbox to\hsize{\vrule\@width 0.4pt\@height#1\hfil
\vrule\@width 0.4pt\@height#1}\hrule\@height 0.4pt\@width\hsize}}

\slugcomment{The Astrophysical Journal, {\bf 750}, 140 (Received 2012 January 31,
  accepted 2012 March 10)}

\shorttitle{The Fate of Intermediate-Age GCs in NGC~1316 (and other Merger Remnants?)}
\shortauthors{Paul Goudfrooij}

\begin{document}


\title{Sizes, Half-Mass Densities, and Mass Functions of Star Clusters in 
  the Merger Remnant NGC~1316: Clues to the Fate of Second-Generation  
  Globular Clusters\altaffilmark{1}}     


\author{Paul Goudfrooij}
\affil{Space Telescope Science Institute, 3700 San Martin
  Drive, Baltimore, MD 21218}
\email{goudfroo@stsci.edu}


\altaffiltext{1}{Based on observations with the NASA/ESA {\it Hubble
    Space Telescope}, obtained at the Space Telescope Science
  Institute, which is operated by the Association of Universities for
  Research in Astronomy, Inc., under NASA contract NAS5-26555} 




\begin{abstract}
We study mass functions of globular clusters derived from {\it HST/ACS\/}
images of the early-type merger remnant galaxy NGC~1316 which hosts a
significant population of metal-rich globular clusters of intermediate age
($\sim$\,3 Gyr). For the old, metal-poor (`blue') clusters, the peak mass of
the mass function $\cM_{\rm p}$ increases with internal half-mass density
$\rho_{\rm h}$ as $\cM_{\rm p} \propto \rho_{\rm h}^{0.44}$ whereas it stays
approximately constant with galactocentric distance $\Rgal$. The mass
functions of these clusters are consistent with a simple scenario in which
they formed with a Schechter initial mass function and evolved subsequently by
internal two-body relaxation. For the intermediate-age population of
metal-rich (`red') clusters, the faint end of the previously reported
power-law luminosity function of the clusters with $\Rgal > 9$ kpc is due to
many of those clusters having radii {\it larger than the theoretical maximum
  value imposed by the tidal field of NGC~1316\/} at their $\Rgal$. This
renders disruption by two-body relaxation ineffective. Only a few such diffuse
clusters are found in the inner regions of NGC~1316. Completeness tests
indicate that this is a physical effect. Using comparisons with star clusters
in other galaxies and cluster disruption calculations using published models,
we hypothesize that most red clusters in the low-$\rho_{\rm h}$ tail of the
initial distribution have already been destroyed in the inner regions of
NGC~1316 by tidal shocking, and that several remaining low-$\rho_{\rm h}$ 
clusters will evolve dynamically to become similar to `faint fuzzies' that
exist in several lenticular galaxies. Finally, we discuss the nature of
diffuse red clusters in early-type galaxies. 
\end{abstract} 




\section{Introduction}              \label{s:intro}
Globular star clusters (GCs) have long been recognized as important
laboratories in the study of the formation and evolution of galaxies for a
variety of reasons. Studies of star formation within molecular clouds
using infrared observations have shown that most, if not all, stars form in
clusters with initial masses $\cM_{\rm cl,\,0}$ in the range $10^2 - 10^8$
\Msun\ \citep[e.g.,][and references therein]{ladlad03,pz+10}. While most star
clusters with $\cM_{\rm cl,\,0} \la 10^4$ \Msun\ are thought to disperse into
the field population of galaxies within a few Gyr by a variety of disruption
processes, the surviving massive GCs constitute luminous compact sources
that can be easily observed out to distances of several tens of
megaparsecs. Furthermore, star clusters represent very good approximations of a
``simple stellar population'' (hereafter SSP), i.e., a coeval population of
stars with a single metallicity, whereas the diffuse light of galaxies is
typically composed of a mixture of populations. 
Thus, star clusters represent invaluable probes of the SFR and chemical
enrichment occurring during the main star formation epochs within a
galaxy's assembly history.   

As to the class of ``normal'' giant early-type galaxies 
(E and S0 galaxies, i.e., galaxies whose light is dominated by a
``bulge'' component), deep imaging studies with the {\it Hubble Space
  Telescope (HST)\/} revealed that such galaxies typically contain rich GC
systems with bimodal color distributions
\citep{kunwhi01,lars+01,peng+06a}. Follow-up spectroscopy with 8-m
class telescopes showed that both ``blue'' and ``red'' GC subpopulations are
nearly universally old, with ages $\ga$\,8 Gyr
\citep[e.g.][]{forb+01,puzi+05,puzi+06,wood+09}, similar to the case of GCs in
our Galaxy \citep{mari+09}. This implies that the color bimodality is due
mainly to differences in metallicity. The colors and spatial distribution of
the blue GCs are usually consistent with those of metal-poor halo GCs in our
Galaxy and M31, while red GCs have colors and spatial distributions that are
similar to those of the ``bulge'' light of their host galaxies
\citep{geis+96,forb+97,rhozep01,goud+01b,jord+04,peng+06a,goud+07}. Thus,  
the nature of the red metal-rich GCs is likely to hold important clues to the
star formation history of their host galaxies. 

The star formation activity associated with the formation of the bulge
component of giant early-type galaxies is generally thought to be dominated by
vigorous starbursts triggered by mergers and interactions of gas-rich galaxies 
\citep[e.g.,][]{whifre91,baug+96,mihher96,cole+00,some+01}. Indeed, deep 
images taken with the {\it HST\/} have shown that nearby galaxy mergers and
young merger remnant galaxies host very rich populations of young star
clusters with masses reaching 10$^7$ \Msun\ and beyond
\citep[e.g.,][]{holt+92,schw+96,mill+97,whit+10}. The ages
and masses of these young clusters predicted from their colors and
luminosities were confirmed by spectroscopy
\citep{zepf+95,schsei98,mara+04,bast+06}. Their metallicities tend to be
near-solar, as expected for clusters formed out of enriched gas in spiral
galaxies. As to the question whether or not these young metal-rich GC
populations can survive several Gyr of stellar and/or dynamical evolution
\citep[this was brought up by][]{brod+98}, studies of GCs in early-type
merger remnants of intermediate age (2\,--\,4 Gyr) have shown that such
metal-rich GC populations do survive at least that long with relatively minor
changes in their mass function down to the detection limit 
\citep{goud+01a,goud+01b,whit+02,goud+04,goud+07}. 
It seems natural to interpret these data in the sense that the
bulges of normal giant early-type galaxies and their metal-rich GCs formed at
high redshift in a way similar to that observed in gas-rich galaxy mergers
today, as originally predicted by \citet{schw87} and \citet{ashzep92}. 

A significant difference between young GC systems in mergers and ancient
GC systems is the shape of their respective luminosity functions (LFs). Young
GC systems in mergers and young merger remnants have LFs consistent with 
power laws: $\phi(L) \propto L^\alpha$ with $\alpha \simeq -2$
\citep[e.g.,][]{schw+96,mill+97,zepf+99,zhafal99,fall+09}, without a sign of a
peak or ``turnover'' down to the detection limit of the data. Perhaps not
surprisingly, this distribution is consistent with the mass function of giant
molecular clouds in star forming regions in our Galaxy \citep{elmefr97}. In
contrast, the LFs of GC systems in ``normal'' galaxies are well described by
Gaussians in magnitude space (or lognormal luminosity distributions),
peaking at $M_V^0 \simeq -7.4$ mag \citep[e.g.,][]{harr01,jord+07},
corresponding roughly to a ``turnover'' mass 
$\cM_{\rm TO} \simeq 1.5\times10^5$ \Msun. If one accepts the view that ancient GCs
in normal galaxies formed in the early universe through essentially the
same processes as star clusters in galaxy mergers today, then large numbers of
low-mass clusters must have been destroyed or disrupted in the mean
time. Recent studies of dynamical evolution of GCs through mechanisms
acting on long time scales (notably two-body relaxation and tidal shocking)
have shown that low-mass clusters disrupt first as galaxies age, which can
evolve a power-law mass function into lognormal mass functions in a few Gyr 
(e.g., \citealt{falree77,falzha01,prigne08}, but see
\citealt{vesp01}). However, there is some debate as to the impact of the tidal
field of the host galaxy on the way GC mass functions (GCMFs) are shaped over
time. One school of thought is that long-term cluster disruption is dominated
by {\it internal\/} two-body relaxation which is essentially independent of
the tidal field of the galaxy. This view is supported by properties
of the GCMFs of the Milky Way and the Sombrero galaxy
\citep{chan+07,mclfal08}. Another view is that the strength of the tidal field
must be a significant factor in terms of the impact of two-body
relaxation 
\citep[e.g.][]{baumak03,giebau08,baum+08}. This in turn seems to be supported
by the finding that the turnover luminosity of GCLFs of early-type
galaxies decreases with decreasing galaxy luminosity \citep{jord+07}. 
Studies of GC systems of intermediate-age (1\,--\,5 Gyr old) merger remnants 
can help clarify the relative importance of different dynamical
effects on GC systems formed during galaxy interactions.  

Several studies have established firmly that the giant galaxy NGC~1316 (=
Fornax A = Arp 154), the dominant galaxy of a subgroup of the Fornax cluster
of galaxies, is a {\it bona fide\/} intermediate-age merger remnant. 
Its outer envelope includes several
nonconcentric arcs, tails, and loops that are most likely remnants of tidal
perturbations, while the inner part of the spheroid is characterized by a
surprisingly high central surface brightness and small effective radius for
the galaxy's luminosity \citep{schw80,schw81,caon+94}. All of these features
indicate that  NGC 1316 is the product of a dissipative merger with incomplete
dynamical relaxation. 
\citet{goud+01a,goud+01b} discovered a significant
population of $\simeq$~3-Gyr-old GCs of near-solar metallicity through a
comparison of {\it BVIJHK\/} colors as well as H$\alpha$ and Ca\,{\sc ii}
triplet line  strengths with SSP model predictions. 
It is classified as a lenticular galaxy in galaxy catalogs: (R$'$)SAB(s)0 in
the RC3 catalog \citep{rc3} and S0$_1$pec in the RSA catalog \citep{rsa}, and
its stellar body is rotationally supported \citep[e.g.,][]{arna+98}. 
In accordance with our earlier work, we adopt a distance of 22.9 Mpc for NGC
1316 in this paper. This distance was derived using known tight
relations between the light curve shape, luminosity, and color of
type Ia supernovae in NGC~1316 \citep[see][and references therein]{goud+01b}. 

In a previous paper \citep{goud+04}, we used deep photometry taken with the
Advanced Camera for Surveys (ACS) aboard {\it HST\/} and showed that the
inner 50\% of the system of intermediate-age GCs of NGC 1316
showed evidence of the presence of a turnover in its LF, while the GCLF of the
outer 50\% was still consistent with a power-law down to the detection
limit. While this provided evidence that long-term dynamical evolution of GCs
formed in mergers of massive gas-rich galaxies can indeed transform MFs of
``young'' clusters to MFs of ancient GC systems, it did not clarify the fate
of the clusters in the outer regions of NGC 1316. A specific issue in that regard
is that GCLFs and GCMFs in the outskirts of ``normal'' galaxies typically show
very similar shapes and turnovers to those in their inner regions, and the
question is whether and how dynamical evolution of the GCs in the outskirts of
NGC~1316 may create a mass function similar to those seen in ``normal'' giant
early-type galaxies over the next $\sim$\,10 Gyr. Using new GC size
measurements, this issue is one of the main topics being addressed in the
current paper. 

The remainder of this paper is organized as follows. After a brief description
of the imaging data in Section \ref{s:data}, the data analysis is presented in
Section \ref{s:anal}. Section~\ref{s:demogr} describes relations between
radii and other properties of the blue and red GC subpopulations in NGC~1316
and presents their mass functions as functions of mass density and
galactocentric radius. In Section~\ref{s:evol_iagc}, we apply various
dynamical evolution model calculations to the red, intermediate-age clusters
and evaluate the properties of their resulting mass functions and their
distribution in the mass-radius plane. A summary and discussion of the
results, including the large number of diffuse red clusters in NGC~1316, 
is provided in Section~\ref{s:disc}.   

\section{Data} \label{s:data}
We use the {\it HST/ACS\/} imaging dataset described previously in
\citet{goud+04}. Briefly, this consists of several wide-field channel (WFC)
images taken with the F435W, F555W, and F814W filters with total
exposure times of 1860 s, 6980 s, and 4860 s, respectively. 
The sample of star cluster candidates in NGC~1316 and its
subdivision in ``old, metal-poor'' (``blue'') versus ``intermediate-age,
metal-rich'' (``red'') clusters was adopted from \citet{goud+04}. For purposes
of dynamical evolution calculations, we assume throughout this paper that the
system of ``red'' GCs in NGC~1316 is dominated by GCs of an age of 3
Gyr, which is the age found for the brightest red GCs using spectroscopy and
multi-color photometry \citep{goud+01a,goud+01b}. While it is 
possible that (one of) the progenitor galaxies possessed older metal-rich GCs
that were ``donated'' to NGC 1316 during the merger, the LF of the red GCs
does not show any sign of the presence of a significant number of ``old'' GCs
such as a turnover near $M_V^0 \simeq -7.4$ mag
\citep[see][]{goud+04}. Instead, the LF of the red GCs continues to rise to
the completeness limit of the data similar to the situation in other merger
remnants. Hence we deem it unlikely that NGC~1316 hosts {\it significant\/}
numbers of old red GCs formerly associated with merger progenitor galaxies.   
 However, the impact of the presence of a small
 fraction of old metal-rich GCs to the results shown in this paper
 will be discussed where relevant.

\section{Analysis}  \label{s:anal}

\subsection{Size Measurements}  \label{s:sizes}

GC structural parameters 
were derived using the {\sc ishape} algorithm
\citep{lars99}, which fits the object's surface brightness profile
with analytical models convolved with a subsampled PSF. Since GCs at the
distance of NGC~1316 are only marginally resolved, reliable size measurements
require a very good knowledge of the ACS PSF at the time of observation. 
While many studies build empirical PSFs from several well-exposed stars in the
field of view for this purpose, the ACS images of NGC~1316
do not contain enough such stars to create a robust set of PSFs. Hence
we have to rely on external PSF libraries. 
To test the influence of the PSF library chosen, we 
performed a comparison between $r_{\rm h}$ values measured using PSFs created
by means of two methods: {\it (i)\/} the
\anchor{http://www.stsci.edu/software/tinytim/tinytim.html}{{\sc TinyTim}}
package \citep{krihoo04} which takes into account the field-dependent
aberration of the ACS/WFC camera, filter passband effects, charge diffusion
variations, and varying pixel area due to the significant field distortion in
the ACS/WFC field of view. The ten times subsampled Tiny Tim PSF was
evaluated at the position of each point-like object in the image and
convolved with the charge diffusion kernel. {\it (ii)\/} the empirical
grid of ACS PSFs of \citet{andkin06}. Since the latter PSFs are
created for individual ACS {\tt *\_flt.fits} images rather than images
combined with {\sc multidrizzle}, we used Maurizio Paolillo's
\anchor{http://www.na.infn.it/~paolillo/Software.html}{{\it Multiking\/}} 
suite of scripts (which we modified to work on filters other than F606W) to
create a grid of empirical PSFs at the location of each GC 
candidate in blank versions of each individual input {\tt
  \_flt.fits} ACS image. These images were then combined by
MultiDrizzle in the exact same way as the final NGC~1316 image. These
PSFs were then subsampled by a factor 10 in order for them to be used
appropriately within {\sc ishape}, using polynomial interpolation. We
refer to these PSFs as {\it ePSFs\/} in the following discussion. 
These two sets of PSFs were then used within {\sc ishape} to fit
the objects' profiles with King (\citeyear{king62}) model 
concentration parameters $C_K \equiv r_{\rm t}/r_{\rm c}$ (where
$r_{\rm t}$ is the tidal radius and $r_{\rm c}$ is the core radius) of 
5, 15, 30, and 100. GC structural parameters were calculated using the
$C_K$ value that yielded the lowest $\chi^2$. 

The effect of the fitting radius (parameter {\sc fitrad} within {\sc
  ishape}) on the derived cluster radii was assessed by using values
of 4, 6, 10, and 15 pixels. While larger values of {\sc fitrad} are in
principle preferable over smaller ones (especially for more extended
sources), we found that the associated increase in fitting
uncertainties for fainter sources renders this advantage
negligible. Similar to \citet{carhol01}, we find that {\sc fitrad}
$\geq$~6 pixels yield consistent $\rh$ values within the
uncertainties, even for the largest GCs found in this study (which are
mainly the fainter ones, see Sect.~\ref{s:rh_vs_Mv} and beyond). Hence
we adopt {\sc fitrad} = 6 pixels for this study. 
GC half-light radii
$\rh$ were derived from the geometric mean of the FWHM values measured along the
semi-major and semi-minor axes. The measured values of $\rh$ were multiplied
by a factor 4/3 to account for projection effects \citep{spit87}.  In order to
render correct values of signal-to-noise ratio (S/N) for the fits, we added 
the local background level back into the model-subtracted image (using the
{\sc mknoise} task within IRAF) for each object prior to running {\sc ishape}
on it. Objects fit by {\sc ishape} with S/N $\ge 50$ were considered for the
remaining analysis. Several studies found this S/N constraint sufficient to
yield reliable half-light radii when using methods such as {\sc ishape} 
\citep[e.g.,][in preparation]{lars99,harr09,puzi+12}\footnote{
We note that a much higher value of S/N is required to determine King model 
concentration indices robustly for marginally resolved objects \citep[S/N $\ga
150$, see][]{carhol01}. Hence, the best-fit values of $C_K$ determined
here for each GC should typically be regarded as estimates.}.

\begin{figure}[tb]
\centerline{
\psfig{figure=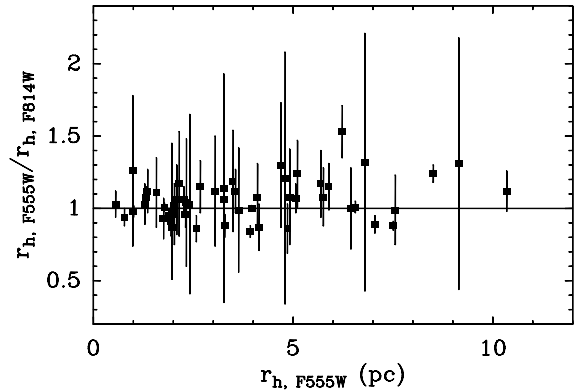,width=7.5cm}
}
\caption{
$r_{\rm h,\,F555W}/r_{\rm h,\,F814W}$ (the ratio of $\rh$ values
measured in the F555W and F814W passbands) vs.\ $\rh$ in 
F555W for all GC candidates with S/N $\geq$ 50 in both
passbands. The solid line indicates a ratio of unity. Error bars are
only shown for $r_{\rm h,\,F555W}/r_{\rm h,\,F814W}$ to avoid
clutter. There does not seem to be a significant systematic difference
between $\rh$ values measured in F555W and F814W.
\label{f:rh_ratioplot}}
\end{figure}

\begin{figure}[tb]
\centerline{
\psfig{figure=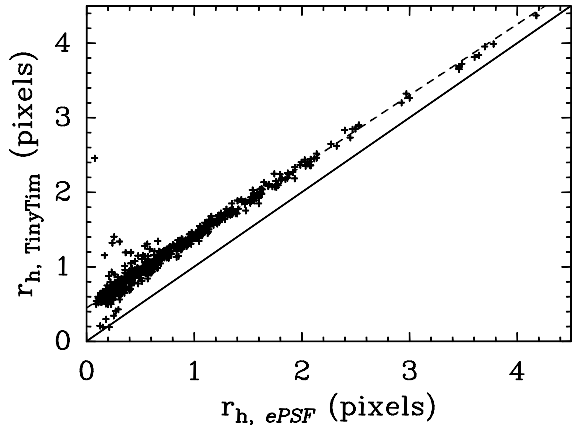,width=7.5cm}
}
\caption{
Comparison between $r_{\rm h}$ values of GC candidates in NGC~1316
derived using PSFs from {\it ePSF\/} (abscissa) and {\sc TinyTim}. The
solid line depicts a 1:1 relation, while the dashed line shows the
best-fit line to the data. 
\label{f:Rhcomp}}
\end{figure}

To test for systematic differences between sizes measured in the 
F555W and F814W images, we plot the ratio $r_{\rm h,\,F555W}/r_{\rm
  h,\,F814W}$ versus $r_{\rm h,\,F555W}$ in
Figure~\ref{f:rh_ratioplot} for all clusters measured with S/N $\ge
50$ in both passbands. Weighting by inverse variance, the weighted mean
value $\left< r_{\rm h,\,F555W}/r_{\rm h,\,F814W} \right> = 0.98 \pm 0.02$. We thus
assume that there is no significant systematic difference between
sizes measured in F555W versus F814W for this dataset. 
In the following, we concentrate on sizes measured in the F555W
image because {\it (i)\/} the F555W image reaches the highest S/N ratio
for cluster candidates in the color parameter space of interest and
{\it (ii)\/} F555W images were taken using the highest number of
distinct spatial offsets (6 ``dithers'') between individual images
which renders the highest spatial resolution after image combination
using the {\sc multidrizzle} task available within PyRAF/STSDAS.
The constraint of S/N $\geq 50$ in the F555W image 
yielded a total of 424 GCs, 212 of which are ``red'' and 212 are ``blue''.  
Fig.~\ref{f:Rhcomp} depicts a comparison between the $r_{\rm h}$
values derived from the best-fit {\sc TinyTim} PSFs and those derived from the
{\it ePSF}s. The $r_{\rm h}$ values derived from the {\it ePSFs\/} are
systematically smaller than those derived from the {\sc TinyTim} PSFs by
$0.45 \pm 0.05$ pixel. Effects of this order of magnitude (differences of 0.4
-- 0.5 pixel) were also seen in previous studies using {\it HST/ACS\/}
images that compared GC sizes derived with {\sc TinyTim} PSFs versus
empirical PSFs \citep[e.g.,][]{spit+06,geor+09a}. This discrepancy is
likely due to effects that are not incorporated in {\sc TinyTim} such
as telescope focus changes among and during {\it HST\/} orbits
(``breathing'') as well as slight broadening of the effective PSF due
to the effect of image combination. With this in mind, we adopt the
structural parameters of GCs measured using the {\it ePSF}s. 

\begin{figure}[tb]
\centerline{
\psfig{figure=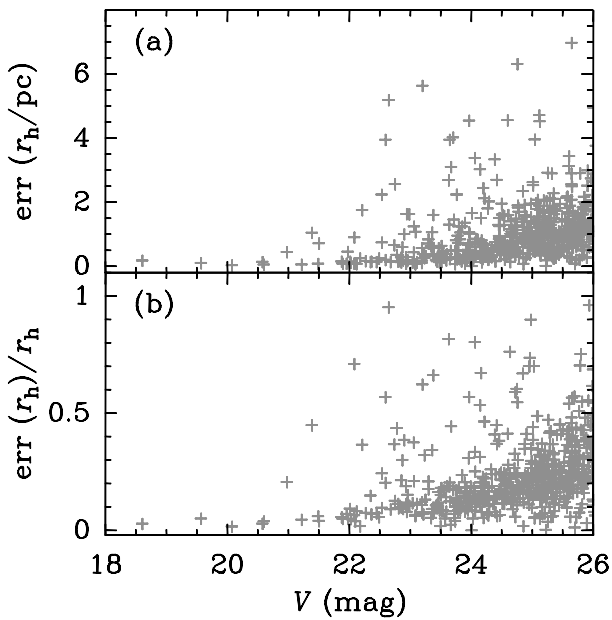,width=8.cm}
}
\caption{
{\it Panel (a)}: err\,($\rh$) (the standard deviation of $\rh$) versus $V$
magnitude for GC candidates with size measurements from the F555W image. {\it
  Panel (b)}: err\,($\rh$)/$\rh$ versus $V$ magnitude for the same
objects. 
\label{f:errorplot}}
\end{figure}

Finally, Figure~\ref{f:errorplot} depicts the formal standard deviation of
$\rh$ as a function of magnitude calculated from the {\sc ishape}
output, both in an absolute sense and relative to the values of
$\rh$. Standard deviations were calculated using {\sc ishape}
parameter {\sc correrr = yes}, meaning that correlated errors between
different {\sc ishape} parameters were taken into
account. Table~\ref{t:sizetab} lists magnitudes, colors, $\rh$ values,
and $C_K$ values for all GC candidates used in this paper. 

\begin{table*}[tbh]
\begin{center}
\caption{Photometry, astrometry, and sizes of GC candidates in NGC~1316\tablenotemark{a}.}
 \label{t:sizetab}
\begin{tabular}{@{}cccccccl@{}}
\multicolumn{3}{c}{~} \\ [-2.5ex]   
 \tableline 
\multicolumn{3}{c}{~} \\ [-2.ex]   
\tableline
\multicolumn{3}{c}{~} \\ [-1.8ex] 
ID & ID$_{\rm 2001b}$ & RA & DEC & $V$ & $V\!-\!I$ & $\rh$ & $C_K$ \\
\multicolumn{1}{c}{(1)} & \multicolumn{1}{c}{(2)} & \multicolumn{1}{c}{(3)}  & (4) & 
 (5) & (6) & (7) & (8) \\ [0.5ex] \tableline 
\multicolumn{3}{c}{~} \\ [-1.5ex]              
\colzero     1\cola  114 & 50.6770102\colb  $-$37.2111334\colc  18.607 $\pm$     0.002\cole     1.093  $\pm$  0.002\colg   6.23 $\pm$ 0.18\coli 30\eol
\colzero     2\cola      & 50.6781450\colb  $-$37.2045648\colc  19.570 $\pm$     0.002\cole     1.013  $\pm$  0.002\colg   2.03 $\pm$ 0.10\coli 30\eol
\colzero     3\cola  210 & 50.6585913\colb  $-$37.2186257\colc  20.078 $\pm$     0.002\cole     1.003  $\pm$  0.002\colg   1.70 $\pm$ 0.03\coli 30\eol
\colzero     4\cola  110 & 50.6522255\colb  $-$37.1822058\colc  20.578 $\pm$     0.004\cole     1.044  $\pm$  0.006\colg   4.74 $\pm$ 0.14\coli 30\eol
\colzero     5\cola      & 50.6923963\colb  $-$37.2013999\colc  20.601 $\pm$     0.004\cole     1.055  $\pm$  0.006\colg   1.40 $\pm$ 0.05\coli 30\eol 
\multicolumn{3}{c}{~} \\ [-2.ex] \tableline
\end{tabular}
\tablecomments{
Column (1): Object ID (sorted by $V$ magnitude). 
(2): Object ID in \citet{goud+01b}. 
(3) and (4): Right ascension (RA) and Declination (DEC) (both in degrees) in J2000 equinox. 
 For reference, the coordinates of the center of NGC 1316 on the F555W ACS image is 
 (RA, DEC) = (50.6739027, $-$37.2079400). 
(5): $V$ mag (Vega-based). 
(6): $V\!-\!I$ color in mag.
(7): Measured half-light radius in pc. 
(8): King concentration index yielding the lowest $\chi^2$. 
}   
\tablenotetext{a}{
Table \ref{t:sizetab} is published in its entirety in the electronic edition of 
{\it The Astrophysical Journal}. A portion is shown here for guidance regarding its form and content.}
\end{center}
\end{table*}


\subsection{Impact of Varying Cluster Size to Observables} \label{s:sizecorr}

\subsubsection{Aperture Corrections}

\begin{figure}[tb]
\centerline{
\psfig{figure=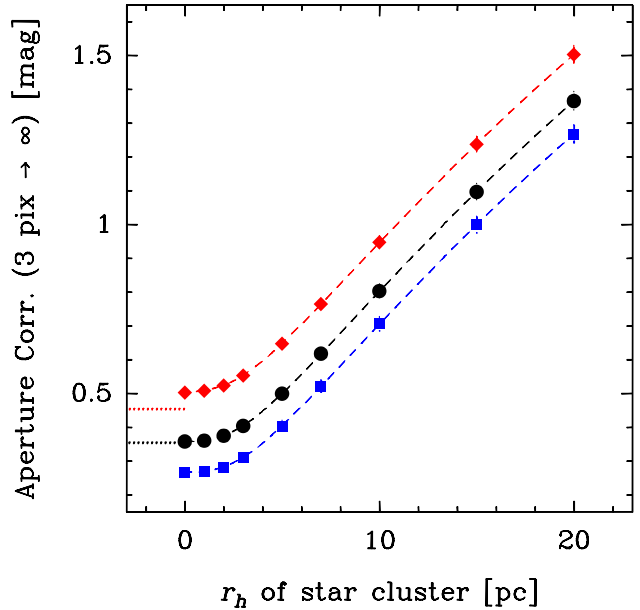,width=8.3cm}
}
\caption{Aperture correction from a measurement radius of 3 pixels to
  ``infinity'' as a function of star cluster half-light radius $r_{\rm
    h}$ at the distance of NGC~1316. Squares represent F435W data, circles
  represent F555W data, and diamonds represent F814W data. For visualization
  purposes, the F555W and F814W data are offset from the F435W data in the Y
  direction by +0.1 and +0.2 mag, respectively. These offsets are indicated by
  dotted lines near the bottom left of the plot. Dashed lines indicate
  polynomial fits to the data. Note the significant increase of aperture
  corrections for clusters with $r_{\rm h} \ga 4$ pc. See \S\,\ref{s:sizecorr}.1.  
\label{f:apercorr}}
\end{figure}

The effect of varying GC sizes to photometric aperture corrections at the
distance of NGC~1316 was determined by using the
\anchor{http://www.na.infn.it/~paolillo/Software.html}{{\it Multiking\/}}
package (see \S\,\ref{s:sizes} above) to create a grid of {\it ePSF}-convolved
King (\citeyear{king62}) models located uniformly throughout the ACS/WFC field
of view for all three ACS/WFC filters used in this dataset. In this regard we
used $r_{\rm h}$ values of 0, 1, 2, 3, 5, 7, 10, 15, and 20 pc and King 
concentration parameters $C_K$ = 5, 15, 30, and 100. Aperture
corrections were calculated from an aperture radius of 3 pixels 
(as used by \citealt{goud+04} for example) to 50 pixels. We found that the
effect of  varying $C_K$ to these aperture corrections at a given $r_{\rm h}$
was negligible relative to the effect of varying $r_{\rm h}$ itself. 
The aperture corrections for the various $r_{\rm
  h}$ values are 
plotted in Fig.~\ref{f:apercorr}. Note that the total intensities of
the {\it ePSF'}s are normalized to unity within a radius of 10 ACS/WFC pixels
\citep{andkin06}. Hence, the aperture correction values from a radius of 10
pixels to infinity listed by \citet{siri+05} were added to the measured values
to arrive at the ``final'' aperture correction values 
plotted in Fig.\ \ref{f:apercorr}. Aperture corrections for all
clusters were applied by means of 5th-order polynomial least-square
fits of aperture correction versus $\rh$. These fits are shown in
Fig.~\ref{f:apercorr} as dashed lines.    

\begin{figure}[tb]
\centerline{
\psfig{figure=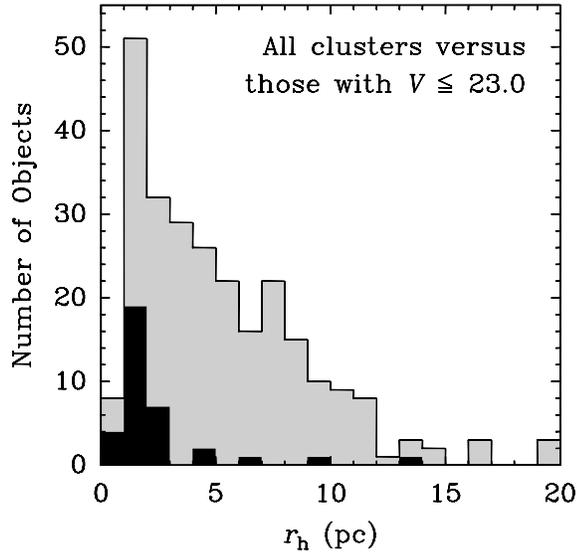,width=7.5cm}
}
\caption{Distributions of half-light radii for all star clusters in NGC~1316
  with reliable size measurements (light-grey histogram) and for the star clusters
  with $V \leq 23$ mag (black histogram). Note that the latter distribution
  misrepresents the former for $\rh \ga 4$ pc. I.e., average aperture corrections
  derived from the brightest clusters yield significantly underestimated
  aperture corrections for clusters with $\rh \ga 4$ pc. 
\label{f:r_h_histfig}}
\end{figure}

Note the significant increase of aperture correction values for
clusters with $\rh \ga 4$ pc, which has a relevant impact on such
clusters' derived luminosities and masses when compared with
``average-sized'' clusters which have $1 \la \rh/\mbox{pc} 
\la 4$. This turns out to have a significant impact for the case of NGC~1316: 
The clusters used by \citet{goud+04} to calculate aperture corrections
were the 35 brightest clusters (with $V \leq 23.0$ mag). While those clusters
obviously render the highest-quality aperture corrections as intended, it
turns out that their $\rh$ distribution is not representative for the cluster
system as a whole. This is illustrated in Fig.~\ref{f:r_h_histfig}: Clusters
with $\rh \ga 4$ pc are quite common among the cluster system as a whole
\citep[see also][]{mast+10} whereas they happen to be strongly
underrepresented among clusters with $V \leq 23.0$ mag.  
Obviously, the measurement of half-light radii can be very important in
photometry studies of star clusters that are marginally resolved \citep[see
also][]{kundu08}. 

\subsubsection{Completeness Corrections}

\begin{figure}[tb]
\centerline{
\psfig{figure=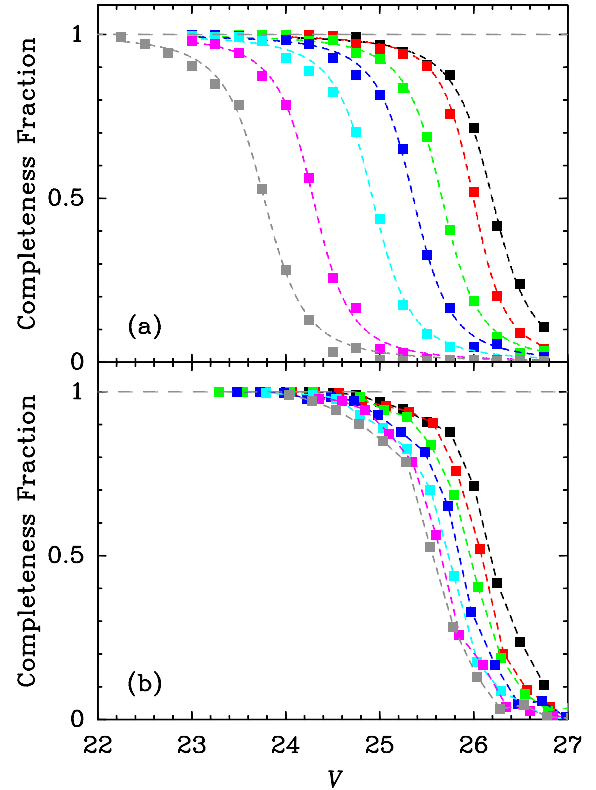,width=7.5cm}
}
\caption{Completeness functions for the ACS photometry of star
  clusters in NGC~1316 as function of {\it total\/} magnitude and $\rh$. A
  typical background level of 200 e$^-$/pix was used for this example. 
  {\it Panel (a)}: Completeness fractions for $\rh$ values (in pc) of 2
  (black symbols), 3 (red), 5 (green), 7 (blue), 10 (cyan), 15 (magenta),
  and 20 (grey). The dashed lines depict the completeness function 
  (equation \ref{eq:completeness}) fit to the data. 
  {\it Panel (b)}: Same data as in panel (a), but now the magnitudes for
  $\rh > 2$ pc have been shifted by the difference in aperture corrections
  between that for the $\rh$ value in question and that for $\rh$ = 2
  pc. See discussion in Sect.\ \ref{s:sizecorr}.2. 
\label{f:compcorr}}
\end{figure}

Given the varying cluster sizes and their possible systematics with
galactocentric distance, we investigated the influence of varying cluster
sizes to completeness corrections by repeating the artificial object tests
done in \citet{goud+04}, but now using the
\anchor{http://www.na.infn.it/~paolillo/Software.html}{{\it Multiking\/}}
package to add simulated GCs with $\rh$ = 1, 2, 3, 5, 7, 10, 15, and 20 pc
and $C_K = 30$ to the final NGC~1316 images in the F555W and F814W
filters. The background levels and (total) magnitude intervals employed for
the completeness tests were identical to those used in \citet{goud+04}, as 
were the parameters used for the {\sc daofind} task within IRAF. The
resulting completeness functions and their dependence on GC size are
illustrated in Fig.\ \ref{f:compcorr} for a typical background level of
200 e$^-$/pix per individual 1100 s exposure.  

Completeness values were parameterized by fitting the measured
completeness fractions $f(V)$ by the following function:
\begin{equation}
f(V) = \frac{1}{2} \left( 1-\frac{\alpha_{\rm c}\,(V-V_{\rm
      lim})}{\sqrt{1\,+\,\alpha_{\rm c}^2 \,(V-V_{\rm lim})^2}}\,\right)\mbox{,} 
\label{eq:completeness}
\end{equation}
where $V_{\rm lim}$ is the 50\% completeness limit in $V$ and the
parameter $\alpha_{\rm c}$ is a measure of the steepness of the completeness
function near $V_{\rm lim}$. Panel (a) of Fig.\ \ref{f:compcorr} depicts
the curves $f(V)$ that were fit to the simulated data of the various $\rh$
values mentioned above, using a least-squares fitting routine to determine
$\alpha_{\rm c}$ and $V_{\rm lim}$. These fits were performed for 5 background
levels encompassing the observed values, equally spaced in
log\,(background). Completeness fractions of all individual clusters were then
determined from their $V$ magnitudes, background 
levels, and $\rh$ values. Values for $\alpha_{\rm c}$ and $V_{\rm lim}$ in the
function $f(V)$ were determined by means of bilinear interpolation in
log\,(background) versus log\,($\rh$) parameter space. 

As shown by panel (a) of Fig.\ \ref{f:compcorr}, the dependence of the
completeness fraction on half-light radius is significant. In quantitative
terms, $V_{\rm lim}$ ranges from 26.2 for $\rh$ = 2 pc to 23.8 for $\rh$ =
20 pc, roughly by 2.4 mag per dex in $\rh$. However, when comparing this
result to the completeness values used in \citet{goud+04}, one should keep
in mind that \citet{goud+04} used a cluster profile based on the brightest 35
clusters for their artificial objects. That cluster profile had a $\rh
\simeq 2$ pc (see Fig.\ \ref{f:r_h_histfig}). 
Since the brightness of artificial objects is scaled by their {\it total\/}
(integrated) magnitude prior to being inserted in an image while 
the photometric measurements are done with an aperture radius of 3
pixels (both here and in \citealt{goud+04}), a proper comparison between the
completeness values determined in this paper versus those in \citet{goud+04}
should take the aperture corrections for clusters of different sizes into
account. Hence we shift the magnitudes in panel (a) of Fig.\ \ref{f:compcorr}
by the size-dependent aperture correction values shown in Fig.\
\ref{f:apercorr}, relative to those appropriate for $\rh$ = 2 pc. The
result is shown in panel (b) of Fig.\ \ref{f:compcorr}. Note that the
``net'' size dependence of completeness values is much reduced relative to
the results shown in panel (a): $V_{\rm lim}$ effectively changes by
only 0.6 mag per dex in $\rh$. 

\subsection{Contamination by Background Galaxies} \label{s:contam}

Contamination by compact background galaxies is always a potentially critical
issue in extragalactic star cluster studies. To address this, we select
images from 4 blank, high-latitude control fields that used ACS/WFC over the
course of 3 HST orbits ($\simeq$ 2.3 hours) with the F555W and F814W filters
from the HST archive. These observations were taken as parallel images in HST
Program GO-11691, and are well suited to the purpose of probing the
background contamination in our cluster sample. Each of these sets of images
were then run through the same image combination, object detection and
selection, and aperture photometry procedures as those employed by
\citet{goud+04} for the NGC~1316 images, with one exception. Since none of
these control fields included large foreground galaxies, the compact-source
photometry in these control fields goes deeper than in the NGC~1316
images. Hence we performed the detection of cluster candidates as if NGC~1316
were included in each blank field. To this end, we used the ``model'' images
generated during the object detection procedure for the NGC~1316 images
\citep[see][]{goud+04} from which we calculated images representing the
Poisson noise associated with the diffuse light of NGC~1316, using the {\sc
  mknoise} program within IRAF. The latter noise images were added to the
blank fields (after scaling by the square root of the exposure times) prior to
running the detection procedure.  

\begin{figure*}[tb]
\centerline{
\psfig{figure=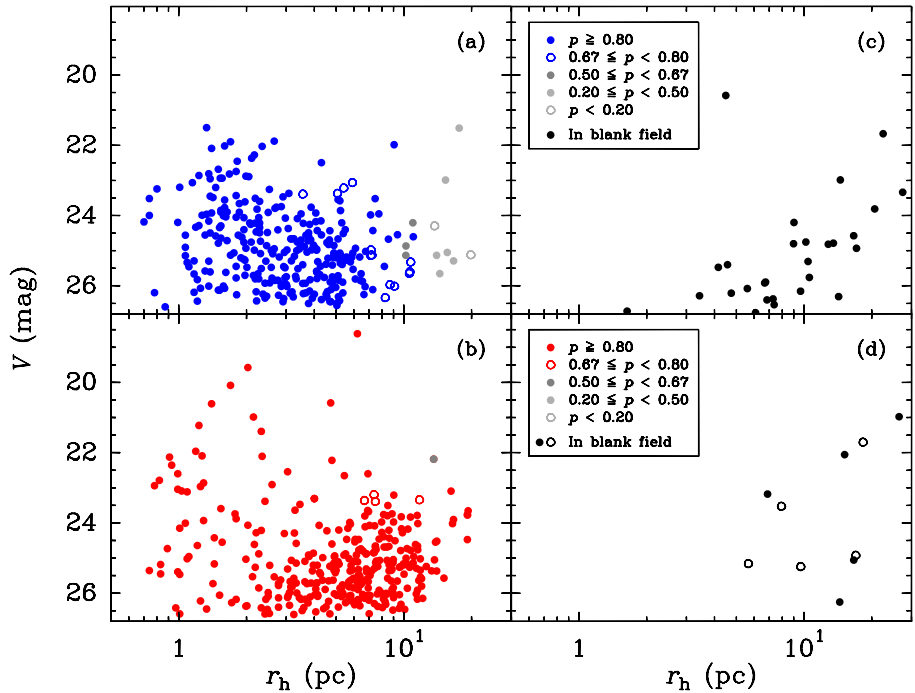,width=12cm}
}
\caption{Size-magnitude diagrams of cluster candidates in NGC 1316 and blank
  fields. Panel (a) shows the ``blue'' (metal-poor) cluster candidates in the NGC
  1316 images while panel (b) does so for the ``red'' (metal-rich) cluster
  candidates. Panel (c) shows a random selection of 1/4 of the ``blue'' objects
  detected in 4 blank sky control fields that were customized for the depth
  reached in the NGC 1316 image. Panel (d) is similar to panel (c), but now
  for 2/4 of the ``red'' objects detected in the 4 control fields (to decrease 
  stochasticity). Symbol colors and types (see boxed-in legends for each pair
  of panels) reflect the probability $p$ that the object in question is
  physically associated with NGC 1316. See discussion in Sect.\
  \ref{s:contam}.  
\label{f:contam}}
\end{figure*}

The level of background contamination of the ``blue'' and ``red'' cluster
subsamples is illustrated in Figure~\ref{f:contam} which shows size-magnitude
diagrams for the data and control fields. The top panels show ``blue''
candidate clusters while the bottom panels show ``red'' candidate
clusters. The control field photometry tables have been randomly sampled so as
to only plot 1/4 of the objects found in the 4 blank fields. Typically, the
relatively small number of ``cluster candidates'' found in the control fields
(which are most likely background galaxies) are fainter and more extended than
the (far more numerous) cluster candidates in the NGC 1316 images. 

To evaluate the probability that a given cluster candidate in the NGC 1316
images is physically associated with NGC 1316, we apply a two-dimensional
Gaussian smoothing kernel to the data in log\,($\rh$) versus $V$ parameter space
to produce density distributions of the numbers of cluster candidates and
contaminants per unit area using the {\sc kde2d} algorithm
\citep{venrip02}. Probability values $p = p\,(\log(\rh),V)$ for cluster
candidates being physically associated with NGC~1316 are then calculated from
the ratio $F_{\rm N1316}/F_{\rm blank}$ where $F_{\rm N1316}$ and $F_{\rm
  blank}$ are the two-dimensional density distributions of cluster candidates
in NGC 1316 and in the blank fields, respectively. For the remainder of this
paper, we define ``clusters'' as objects with $p \geq 0.67$, i.e., objects
with $F_{\rm N1316} \geq 2 \times F_{\rm blank}$.  

\section{Size Demography of the Cluster Subpopulations} \label{s:demogr}

In this Section we illustrate the size-related properties of the metal-poor
(blue) and metal-rich (red) cluster subpopulations in NGC~1316 and attempt to
put them in context by comparing them to clusters in our Galaxy, younger
merger remnant galaxies, and ``normal'', old giant early-type galaxies.  

\subsection{Dependence on Galactocentric Radius} \label{s:rh_vs_Rgal}

\begin{figure*}[tb]
\centerline{
\begin{minipage}[t]{9cm}
\psfig{figure=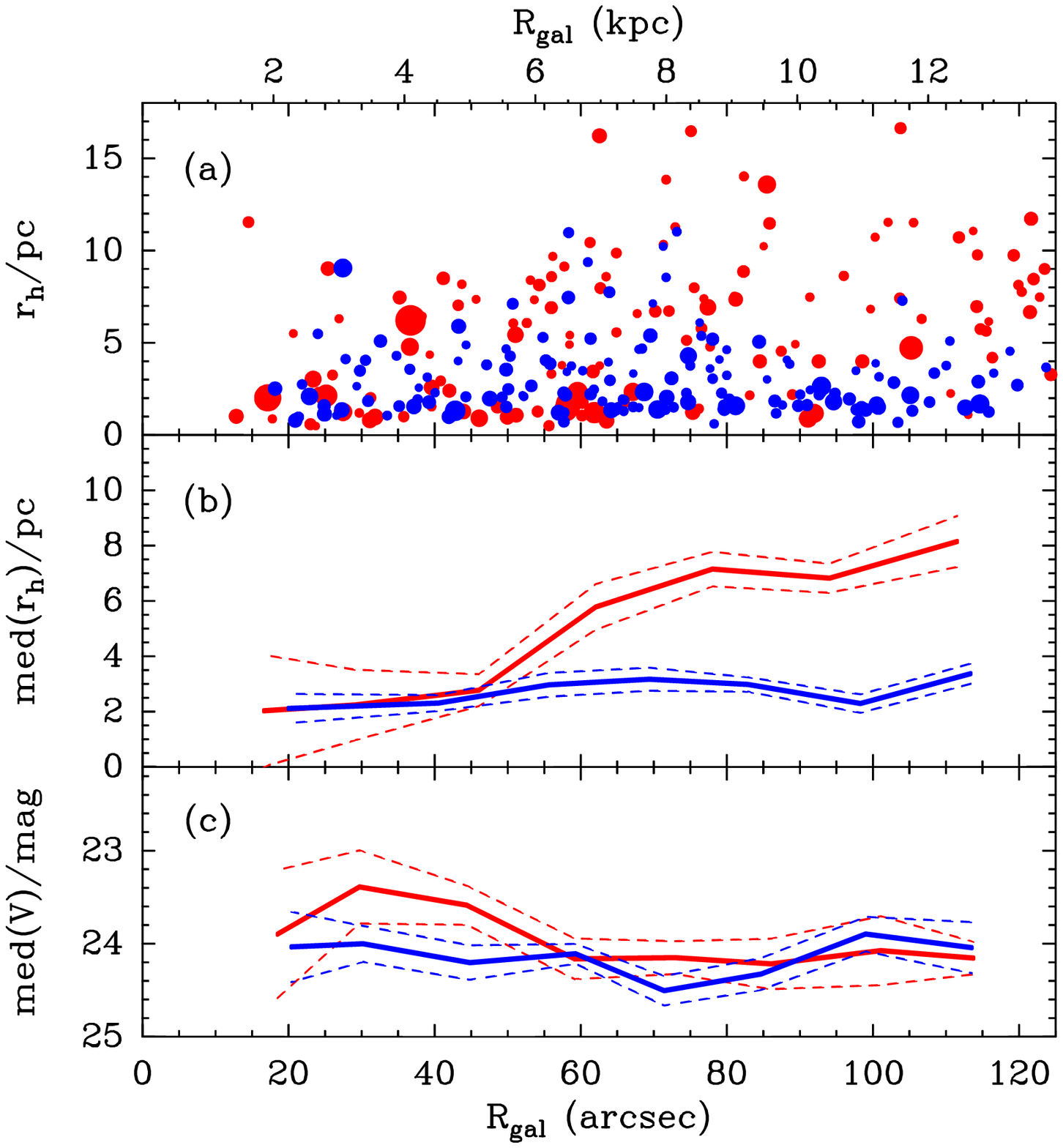,width=9cm}
\end{minipage}
\hspace*{-2.5mm}
\begin{minipage}[t]{4.17cm}
\vbox{
\psfig{figure=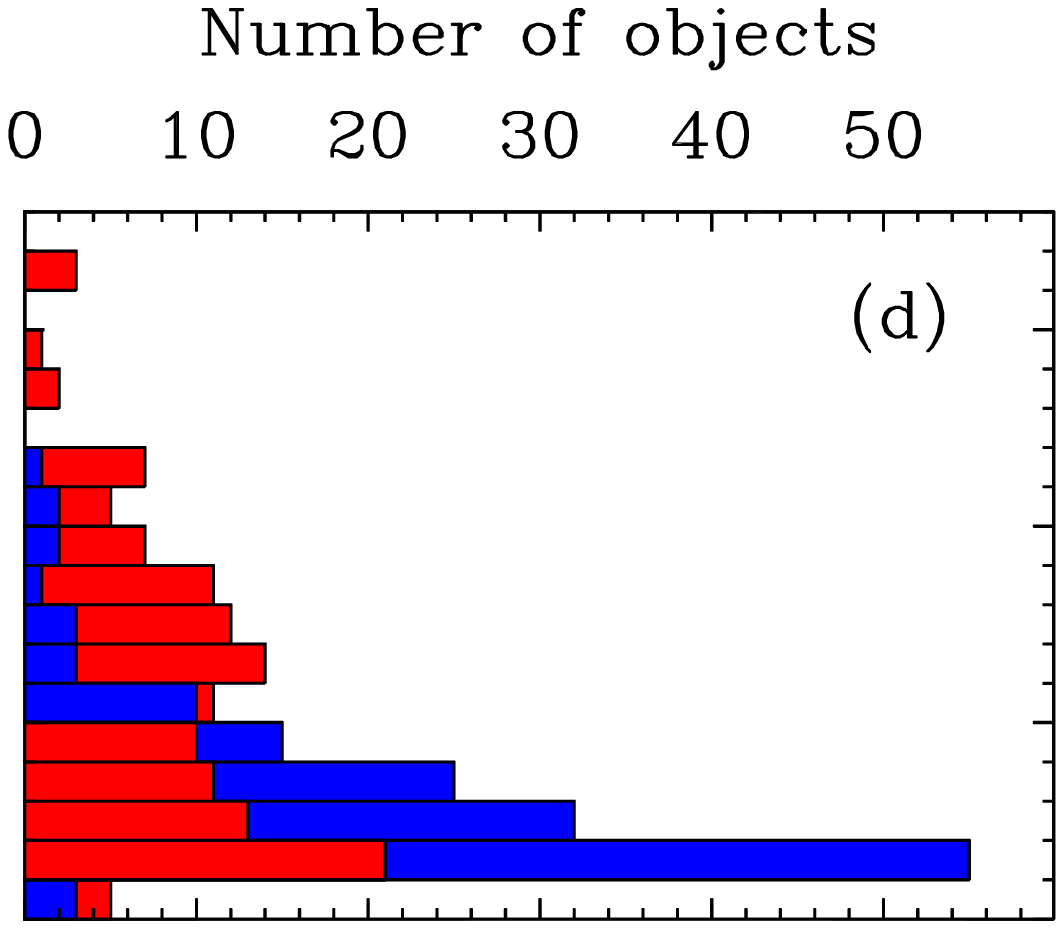,width=4.17cm,angle=-90}
\vspace*{6cm}
}
\end{minipage}
}
\caption{{\it Panel (a)}: GC half-light radius $\rh$ as function of projected
  galactocentric radius $\Rgal$ for individual GCs with $V \leq 25.3$ mag. Blue and
  red symbols or lines represent metal-poor and metal-rich GCs, respectively,
  for {\it all\/} panels. The size of circles scales logarithmically with
  the GCs' $V$-band luminosity.  
  {\it Panel (b)}: Same as panel (a), but now solid lines represent
  running median values of $\rh$. Dashed lines mark the $\pm
  1\sigma$ error of the mean.  
  {\it Panel (c)}: Solid lines represent running median values of $V$
  magnitude as function of $\Rgal$. Dashed lines mark the $\pm
  1\sigma$ error of the mean.
  {\it Panel (d)}: Distribution of $\rh$ values plotted as histograms. 
  See discussion in Sect.\ \ref{s:rh_vs_Rgal}.
\label{f:rh_vs_Rgal}}
\end{figure*}

The top and middle panels of Fig.~\ref{f:rh_vs_Rgal} plot $\rh$ versus
projected galactocentric radius $\Rgal$ for blue and red clusters in
NGC~1316. 
To avoid biases related to varying incompleteness as functions of
$\rh$ and $\Rgal$, only clusters brighter than $V = 25.3$ mag are considered (cf.\
panel b of Fig.\ \ref{f:compcorr}). 
In panel (b) of Fig.\ \ref{f:rh_vs_Rgal} we plot running medians of the $\rh$
distribution as function on $\Rgal$. As can be seen, median sizes of the blue
GCs are  $\widetilde{\rh} \sim$\,2.5\,--\,3.5 pc with only a slight
dependence on $\Rgal$. This is similar to the situation seen for
metal-poor GCs in ``normal'' early-type galaxies
\citep[e.g.,][]{spit+06,harr09,madr+09}. However, the situation for the red
GCs in NGC~1316 is markedly different from that in ``normal'' ellipticals. In
NGC~1316, the red GCs within $\sim$\,5 kpc from the 
galaxy center have sizes similar to those of the blue GCs. Outside $R_{\rm
  gal}$ = 5 kpc however, the median size of the red GCs increases steadily
with increasing galactocentric radius, reaching $\widetilde{\rh} \approx
7.5$ pc at $\Rgal \sim$ 12 kpc. In stark contrast, sizes of red GCs in
``normal'' ellipticals are typically $\sim$\,20\% {\it smaller\/} than blue
GCs \citep[e.g.,][]{kunwhi98,puzi+99,jord+05,mast+10}. The size ratio of blue 
versus red GCs in ``normal'' ellipticals does not change significantly with
$\Rgal$ \citep[][in preparation]{paol+11,puzi+12}.  

To put this result in context, recall that the bulk of red GCs in NGC~1316 most
likely represent an intermediate-age population \citep[age $\sim$\,3 Gyr,
cf.][]{goud+01a,goud+01b,bast+06}. Assuming that to be the case, long-term
cluster disruption mechanisms such as two-body relaxation and tidal shocks
will have been active for {\it much\/} (some 7\,--\,10 Gyr) less long than for 
``old'' red GCs in ``normal'' ellipticals. In particular, late-epoch cluster
contraction due to accumulative effects of two-body relaxation 
\citep[e.g.,][]{mack+08,heggie08,vesp+09} will not yet have taken place. As
such, one might expect the size demography of the red GCs in NGC~1316 to bear 
similarities to that of younger cluster systems 
in otherwise similar environments.
While size measurements of young GCs in massive 
merger remnant
galaxies are still quite sparse in the literature, the available data
do show strong similarities to the red GCs in NGC~1316. 
\citet{tran+07} measured half-light radii of 10\,--\,20 pc for young GCs in the
outskirts of NGC~3256, a remnant of a recent galaxy merger which
\citet{toom77} placed in the middle of his sequence of disk galaxy merger
remnants. These clusters in NGC~3256 have estimated ages of $\simeq 100$ Myr
\citep[see also][]{zepf+99}, and hence they have likely already undergone the cluster
expansion driven by strong mass loss due to supernova type II explosions as
well as stellar evolution in the first few 10$^7$ yr \citep{baum+08,vesp+09},
which renders them an appropriate comparison with the intermediate-age red GCs
in NGC~1316.     
Other young cluster systems 
in merger remnants 
with accurate size measurements also have size distributions similar
to that of the red GCs in NGC~1316, cf.\ the  $\sim$\,300\,--\,700 Myr
old clusters in NGC~1275 and 
NGC~3597 \citep{carhol01}, with mean $\rh$ values of 6.2 and 5.4 pc,
respectively, and a set of 8 clusters
covering the age range $10 \la \mbox{Age/Myr} \la 700$ in the Antennae
galaxies, with a mean $\rh$ of 8.0 pc \citep{bast+09}. 
In conclusion, the relatively large sizes seen among the red GCs in NGC~1316
seem consistent with them being 
intermediate-age counterparts of metal-rich clusters seen in younger
merger remnants. 
The observed increase of the median $\rh$ of the red GCs with increasing $\Rgal$
will be discussed below in Sect.\ \ref{s:evol_iagc}. 

\subsection{Luminosity-Radius Relation} \label{s:rh_vs_Mv}

\begin{figure}[tb]
\centerline{
\psfig{figure=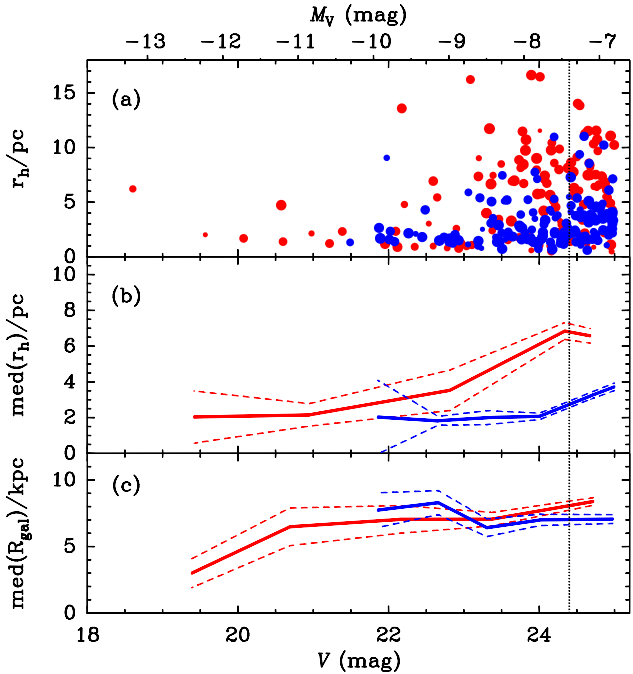,width=8.3cm}
}
\caption{{\it Panel (a)}: GC half-light radius $\rh$ as function of $V$ (or $M_V$, see top
  abscissa) for individual GCs with $V \leq 25.3$ mag. Blue and
  red symbols or lines represent metal-poor and metal-rich GCs, respectively,
  for {\it all\/} panels. The size of the circle scales logarithmically with
  the GCs' galactocentric radius.  
  {\it Panel (b)}: Same as panel (a), but now solid lines represent
  running median values of $\rh$. Dashed lines mark the $\pm
  1\sigma$ error of the mean. 
  {\it Panel (c)}: Solid lines represent running median values of the
  galactocentric radius $\Rgal$ as function of $V$. Dashed lines mark the $\pm
  1\sigma$ error of the mean. The vertical dotted line indicates $M_V = -7.4$,
  the turnover luminosity for ``old'', metal-poor GC systems. 
  See discussion in Sect.\ \ref{s:rh_vs_Mv}.
\label{f:rh_vs_Mv}}
\end{figure}

Panels (a) and (b) of Fig.~\ref{f:rh_vs_Mv} plot $\rh$ versus $M_V$
for the same clusters as those shown in Fig.\ 
\ref{f:rh_vs_Rgal}. With regard to the blue (metal-poor) GCs, previous studies
of sizes of GCs in galaxies found no significant relation between $\rh$ and
luminosity down to the turnover of the GC luminosity function
\citep[e.g.,][]{mcla00,jord+05}. Our results for the blue GCs are consistent with
this. However, our dataset yielded robust $\rh$ values for GCs fainter than
the turnover luminosity for old, metal-poor GCs ($M_V = -7.4$; see Fig.\
\ref{f:rh_vs_Mv}) where we find evidence for an increase of the (median) size
of blue GCs with decreasing luminosity. Panel (c) of Fig.~\ref{f:rh_vs_Mv}
shows that this increase is {\it not\/} due to the increase of $\rh$ with
increasing $\Rgal$ for $\Rgal \ga 11$ kpc for blue GCs shown in
Fig.~\ref{f:rh_vs_Rgal}. Instead, we interpret this effect as being due to dynamical
evolution {\it within\/} GCs, i.e.\ the mass density dependence of cluster
evaporation due to two-body relaxation, as discussed further in Sect.\
\ref{s:survival} below.  

As to the red (metal-rich) GCs, panel (b) of Fig.~\ref{f:rh_vs_Mv} shows an
obvious increase of $\rh$ with decreasing luminosity. However, before trying
to interpret this relation in terms of physical effects, we point out 
that it is likely associated with the correlation between $\rh$
and $\Rgal$ shown in panel (b) of Fig.\ \ref{f:rh_vs_Rgal}. This is
illustrated by panel (c) of Fig.~\ref{f:rh_vs_Mv} which shows that the median
$\Rgal$ also increases significantly with decreasing luminosity. Note that for
GCs with $V \ga 21$ mag, the median $\Rgal$ is in the range where the rate of
increase of $\rh$ with increasing $\Rgal$ is significant (see panel (b) of
Fig.\ \ref{f:rh_vs_Rgal}). These relations are interpreted in terms of
dynamical evolution processes in the next Section.

\subsection{Cluster Distribution in the `Survival Diagram'} \label{s:survival}

\begin{figure*}[ptb]
\centerline{
\psfig{figure=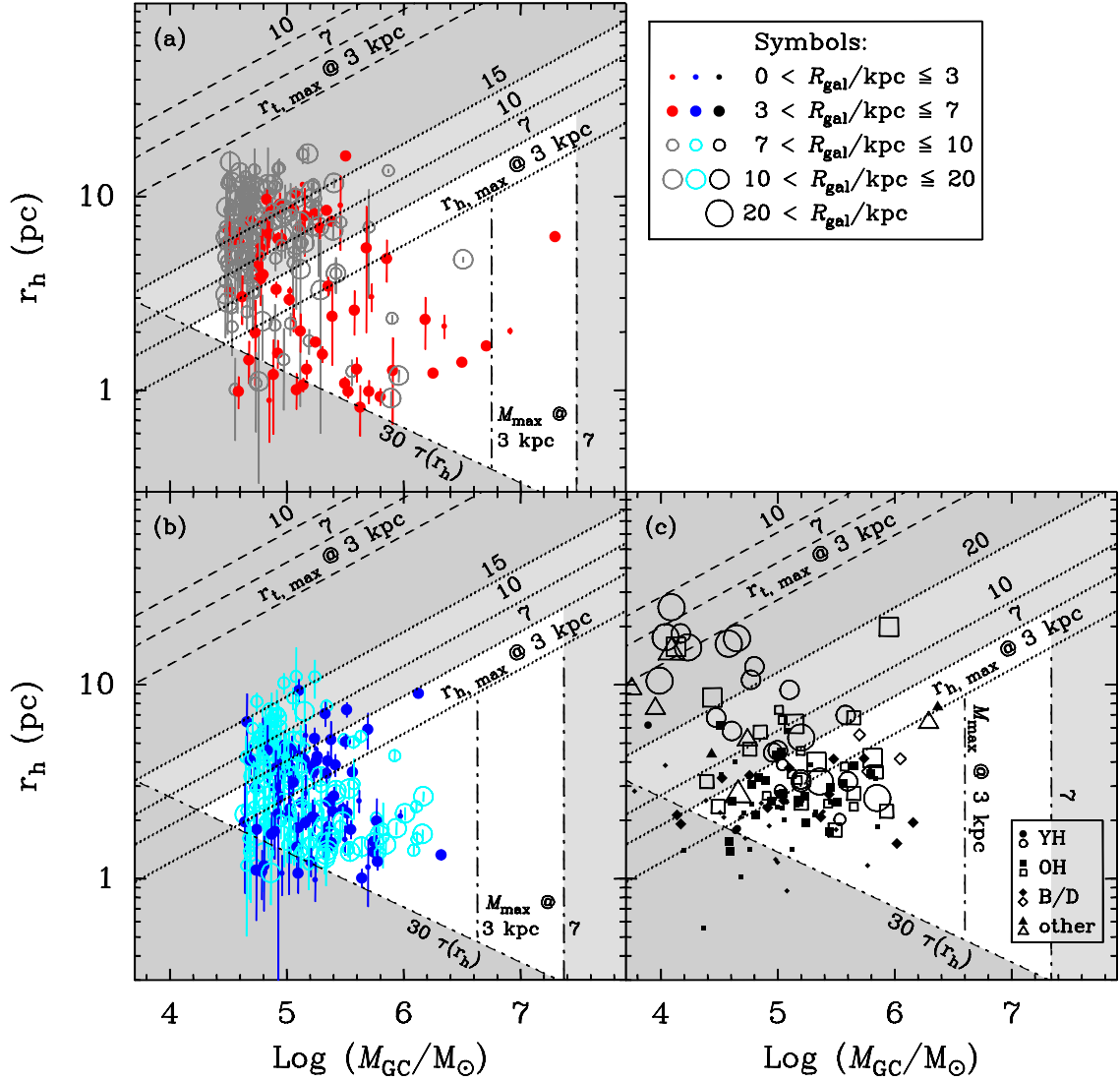,width=15.cm}
}
\caption{Half-light radius versus mass (`survival diagram') for GCs in
  NGC~1316 and our Galaxy. Symbol size indicates (projected) galactocentric
  radius as shown in the legend at the top right. Lines in each panel
  represent limits imposed by dynamical evolution effects at the galactocentric
  radii indicated above the line in question: Short-dashed lines indicate
  max.\ tidal radii, dotted lines indicate max.\ half-light radii,
  short-dashed-dotted lines indicate min.\ radii imposed by two-body
  relaxation, and long-dashed-dotted lines indicate max.\ masses imposed by
  dynamical friction. 
  {\it Panel (a)}: Red GCs in NGC~1316 with $V \le 25.3$. 
  {\it Panel (b)}: Blue GCs in NGC~1316 with $V \le 25.3$. 
  {\it Panel (c)}: GCs in our Galaxy. Each subpopulation of Galactic
  GCs is represented by a different symbol, as shown in the
  legend. Abbreviations for Galactic GC subpopulations: OH = old halo;  
  YH = young halo; B/D = bulge/disk.  
  See discussion in Sect.\ \ref{s:survival}.
\label{f:vitalplot1}}
\end{figure*}

The $\rh$ versus cluster mass diagram is a useful tool to illustrate the
`survivability' of star clusters to various internal and external dissolution
mechanisms \citep[e.g.][]{falree77,gneost97,geor+09a}. 
Fig.\ \ref{f:vitalplot1} shows the distribution of GCs in NGC~1316
with size measurements in this `survival diagram', separately for blue
and red GCs in panels (a) and (b). For comparison purposes, we include
the same diagram for the Galactic GCs in panel (c). To convert
luminosities of GCs in NGC~1316 into masses, we adopt $M/L_V$ values
of the SSP synthesis models of \citet{bc03}, using a \citet{chab03} initial
mass function (IMF)\footnote{Use of the \citet{mara05} SSP models with a
  \citet{krou01} IMF yields $M/L_V$ values that are $\sim$\,5\%
  higher; applying the \citet{salp55} IMF yields $M/L_V$ values higher by
  $\sim$\,75\% relative to Chabrier or Kroupa IMFs.}. We adopt age = 3 Gyr and
[Z/H] = 0.0 for the red GCs \citep{goud+01a,goud+01b} and age = 13 Gyr and
[Z/H] = $-$1.6 for the blue GCs, similar to the average properties of
metal-poor GCs in our Galaxy \citep[e.g.][]{harr96}. For the Galactic GCs,
half-light radii and cluster masses were taken from \citet{mclvdm05} when
available, and from \citet{harr96} for the remainder. Symbol sizes scale with
$\Rgal$, as shown in the legend in the top right of the Figure. 

Fig.\ \ref{f:vitalplot1} includes various lines that depict limits in the
mass-radius plane imposed by various dynamical evolution
mechanisms. To indicate limits imposed by the tidal field of NGC~1316,
we calculate maximum tidal cluster radii $r_{\rm t, max}$ by using the
tidal limit for a satellite on a circular orbit:
\begin{equation}
r_{\rm t, max} = \left(\frac{G {\cal{M}}_{\rm cl}}{2 \, V_{\rm circ}^2} \right)^{1/3}\,
 \Rgal^{2/3}
\label{eq:r_t}
\end{equation}
\citep[e.g.,][]{bintre87,baumak03} where $V_{\rm circ}$ is the circular
velocity of the galaxy and $\Rgal$ is the galactocentric distance. For the
latter, we multiply the observed (projected) galactocentric distances by a
factor $2/\sqrt{3}$ to account for an assumed viewing angle of 60$^{\circ}$. 
We use $V_{\rm circ} = 235$ \kms\ for NGC~1316; this represents the rotation velocity
and velocity dispersion values added in quadrature using velocities of planetary
nebulae and GCs \citep{arna+98,goud+01b}. $r_{\rm t, max}$ is shown in
Fig.\ \ref{f:vitalplot1} by short-dashed lines for three values of $\Rgal$: 3,
7, and 10 kpc. To translate $r_{\rm t, max}$ values to observed half-light
radii $r_{\rm h, max}$, we use the value of $\rh/r_{\rm t}$ for King models
with $C_K = 30$, i.e., $\rh/r_{\rm t} = 0.095$\footnote{This choice was made
  for this figure 
  because $C_K = 30$ provides the best fit for the majority of clusters in
  NGC~1316 and in our Galaxy. For reference, $\rh/r_{\rm t}$ values for $C_K$
  = 15 and 100 are 0.136 and 0.051, respectively.}. Limits $r_{\rm h, max}$
are drawn in Fig.~\ref{f:vitalplot1} for $\Rgal$ = 3, 7, 10, and 15 kpc (for
our Galaxy [i.e., panel c], the latter is 20 kpc), using dotted lines. 

Dash-dotted lines represent limits imposed by cluster evaporation
due to two-body relaxation:
\begin{equation}
r_{\rm h, evap}=\left(\frac{t_{\rm elapse} {\rm [Myr]}}{N(t_{\rm rel})}\right)^{2/3}
\left(\frac{0.138}{\sqrt{G}m_\star \, \ln 
\left(\gamma\frac{{\cal{M}}_{\rm cl}}{m_\star}\right)}\right)^{-2/3} {\cal{M}}_{\rm cl}^{-1/3}
\label{eq:t_evap}
\end{equation}
for clusters of {\it initial\/} mass ${\cal{M}}_{\rm cl}$ that
survived a time $t_{\rm elapse}$ after $N(t_{\rm rel})$ initial
relaxation times; $m_\star$ is the average stellar mass in a 
SSP of the appropriate age and metallicity, and $\gamma=0.02$ is a correction
constant taken from cluster simulations \citep{gieheg96}. GCs with initial
$r_{\rm h} < r_{\rm h, evap}$ will have dissolved after a time 
$t_{\rm elapse}$ of dynamical evolution \citep[e.g.][]{falzha01}. To
facilitate the identification of red GCs in NGC~1316 that are likely to
evaporate after 10 additional Gyr, we show $r_{\rm h, evap}$ for $t_{\rm elapse}
= 10$ Gyr, $N(t_{\rm rel}) = 30$ \citep[see][]{gneost97}, and $m_\star =
0.55$ in panel (a) of Fig.\ \ref{f:vitalplot1}. For panels (b) and (c) of
Fig.\ \ref{f:vitalplot1} we use $t_{\rm elapse} = 13$ Gyr, $N(t_{\rm rel}) =
30$, and $m_\star = 0.49$, appropriate for 13 Gyr old populations with [Z/H]
= $-$1.6. Note  however that the GC masses plotted in Fig.\
\ref{f:vitalplot1} are {\it current\/} masses whereas ${\cal{M}}_{\rm cl}$ in
equation (\ref{eq:t_evap}) refers to {\it initial\/} masses. I.e., one should
take into account that GCs in panels (b) and (c) of Fig.\
\ref{f:vitalplot1} have already undergone some $\sim$\,13 Gyr of
dynamical evolution when comparing the cluster data with the lines
that indicate limits in the mass-radius plane. 

Maximum values for cluster mass imposed by dynamical friction in the
galaxy potential (${\cal{M}}_{\rm cl,\,max}$) are estimated by using
equation [7-26] of 
\citet{bintre87}: 
\begin{equation}
t_{\rm df} = \frac{2.64 \times 10^{10}\, {\rm yr}}{\ln \Lambda} 
 \left(\frac{R_{\rm gal, i}}{2\; {\rm kpc}}\right)^2 
 \left(\frac{V_{\rm cl}}{250\;{\rm km\,s}^{-1}}\right) 
 \left(\frac{10^5 \, {\rm M}_{\odot}}{{\cal{M}}_{\rm cl}}\right)
 \mbox{,} 
\label{eq:t_df}
\end{equation}
where $\ln \Lambda$ is the Coulomb logarithm, $R_{\rm gal, i}$ is the
initial galactocentric distance, and $V_{\rm cl}$ is the
velocity of the cluster with respect to the host galaxy. Values of the
latter two parameters are estimated here by assuming $R_{\rm gal, i} =
\Rgal$ and $V_{\rm cl} = V_{\rm circ}$. Values for ${\cal{M}}_{\rm
  cl,\,max}$ are calculated by solving equation (\ref{eq:t_df}) for
${\cal{M}}_{\rm cl}$ using $t_{\rm df}$ = 10 Gyr for the red GCs in
NGC~1316, and $t_{\rm df}$ = 13 Gyr for the blue GCs in NGC~1316 and
the Galactic GCs. 

Overall, the distribution of blue GCs in NGC~1316 in the mass-radius
plane is very similar to that of (metal-poor) halo GCs in our Galaxy. This is
encouraging, since blue GCs in galaxies are commonly thought to represent the
counterpart of halo GCs in our Galaxy. Note that the tail of the distribution
of Galactic GCs towards low masses and large $\rh$ values is not accessible in
the NGC~1316 ACS data given its detection limit and limited radial
coverage.  

\begin{figure}[tb]
\centerline{
\psfig{figure=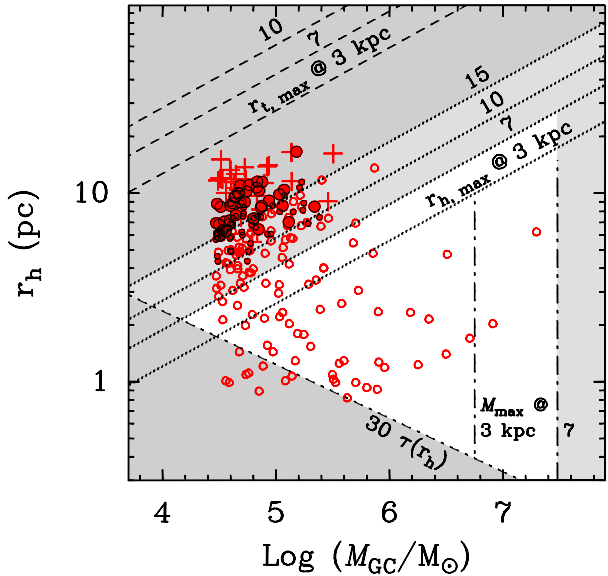,width=8.cm}
}
\caption{Similar to panel (a) of Fig.\ \ref{f:vitalplot1}, except that
  different symbols now indicate different values of the ratio $\rh/r_{\rm h,
    max}$. Open circles indicate GCs with $\rh/r_{\rm h, max} \leq
  1.0$, small filled circles outlined in black indicate GCs with $1.0 <
  \rh/r_{\rm h, max} \leq 1.5$, large filled circles outlined in black
  indicate GCs with $1.5 < \rh/r_{\rm h, max} \leq 2.0$, and pluses
  indicate GCs with $\rh/r_{\rm h, max} > 2.0$. 
  See discussion in Sect.\ \ref{s:survival}.
\label{f:vitalplot2}}
\end{figure}

As to the distribution of red GCs in Fig.\ \ref{f:vitalplot1}, it can be seen
that the majority of the clusters are located in ``healthy'' regions for a
3-Gyr old population, taking their galactocentric radii into
account. Furthermore, there are only a dozen or so GCs that are expected to 
evaporate completely in the next 10 Gyr due to two-body evaporation, and the
effect of dynamical friction is expected to be negligible. However, several
GCs in the top left region of panel (a) of Fig.\ \ref{f:vitalplot1} 
have $\rh$ values that are larger than the $r_{\rm h, max}$ value for their
$\Rgal$. This is illustrated in Fig.\ \ref{f:vitalplot2} which shows
the same diagram as panel (a) of Fig.\ \ref{f:vitalplot1} except that
different symbols now indicate the value of the ratio $\rh/r_{\rm h, max}$. 
It seems likely that many of the red GCs with $\rh/r_{\rm h, max} > 1$ will
experience significant mass loss during ten additional Gyr of tidal shocking. The 
impact of dynamical evolution mechanisms on the distribution of red GCs in the
`survival diagram' and their mass function will be discussed in
Section~\ref{s:evol_iagc}. 

\subsection{Present-Day GC Mass Functions} \label{s:massfunc}

Observations of young star clusters in merging galaxies and young merger
remnants show that the number of clusters per unit mass are well described by
a power law, $dN/d\cM \propto \cM^{\alpha}$ with $\alpha \simeq -2$
\citep[e.g.,][]{fall+09}. 
In the case of ancient GCs in massive ``normal'' galaxies with hundreds or
even thousands of GCs, the mass function decreases more rapidly than a power
law for masses larger than $\sim 3 \times 10^5$ M$_{\odot}$
\citep[see][]{bursmi00}. A function that provides an accurate description of
the shape of mass functions of young clusters as well as the $\sim$\,50\% most
massive GCs in ancient galaxies is the \citet{sche76} function,
$dN/d\cM \propto \cM^{\beta} \exp(-\cM/\cM_{\rm c})$, with power-law component
$\beta \simeq -2$ and exponential cutoff $\cM_{\rm c}$ above some large mass that
might vary among galaxies \citep[e.g.,][and references
therein]{bursmi00,giel+06,jord+07}. This apparent connection between the GC
mass function at high masses and that of young clusters and molecular clouds
has been explained by theoretical models \citep{mclpud96,elmefr97}. 
The main difference between the mass functions of ancient GCs and young
clusters is the fact that the number of clusters per unit mass interval
continues to rise like a power law towards the detection limit for young
clusters, while it stays approximately constant among ancient GCs fainter than
the classic peak magnitude of the GC luminosity function. The latter behavior
is a signature of long-term dynamical evolution driven by two-body relaxation 
\citep{falzha01,jord+07}. 

Another signature of relaxation-driven dynamical evolution is that the
evaporation time scale scales with cluster mass as $\tau_{\rm ev} \propto
\cM_{\rm cl}\,\rho_{\rm h}^{-1/2}$, where $\rho_{\rm h} = 3\cM_{\rm cl}/8\pi
r_{\rm h}^3$ is the mean density inside the cluster's half-mass radius. Under
the common assumption that star clusters conserve their mean half-mass
density while their dynamical evolution is dominated by evaporation
\citep[e.g.,][]{vesp00,vesp01,falzha01}, this means that the peak mass
$\cM_{\rm p}$ of the mass function of ancient cluster systems would scale
with $\rho_{\rm h}$ as $\cM_{\rm p} \propto \rho_{\rm h}^{1/2}$. \citet[][hereafter
MF08]{mclfal08} found roughly this dependence of $\cM_{\rm p}$ on $\rho_{\rm
  h}$ among the old GCs in our Galaxy and argued that this presents evidence
for a scenario that dynamical evolution shaped the mass function of ancient GC
systems and that the dynamical evolution is dominated by two-body relaxation
\citep[see also][showing similar results for the Sombrero galaxy]{chan+07}.

\begin{figure*}[tb]
\centerline{
\psfig{figure=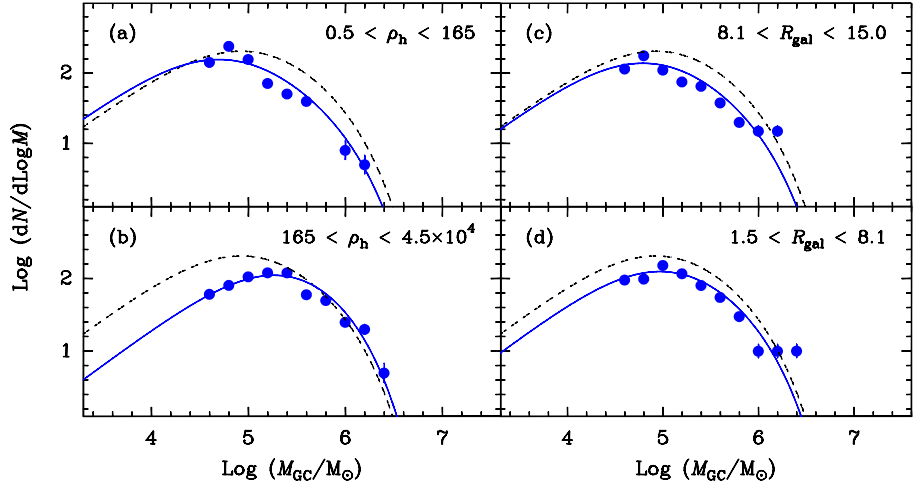,width=12cm}
}
\caption{Mass functions of blue clusters in NGC~1316 with robust size
  measurements. Panels (a) and (b) show data for two ranges of half-mass
  density $\rho_{\rm h}$ in units of M$_{\odot}$ pc$^{-3}$, as indicated in
  the legend. Panels (c) and (d) show data for two ranges of projected
  galactocentric distance $\Rgal$ in kpc, as indicated in the legend. The
  solid curves were computed from equation (\ref{eq:evsch}) by summing over
  mass-loss terms for all individual clusters within the bins of $\rho_{\rm
    h}$ or $\Rgal$ mentioned in each panel. For comparison purposes, the
  dashed curve in each panel was computed from equation (\ref{eq:evsch}) with a
  single term using the median value of $\rho_{\rm h}$ for the full sample of
  212 blue clusters.  
\label{f:MassFunc_Blue}}
\end{figure*}

In the context of these recent studies, we reanalyze the mass functions of the
blue and red GCs in NGC~1316 that have robust size measurements in the
remainder of this Section. 
The completeness-corrected GC mass functions of the blue and red GCs with
robust size measurements are shown in Figs.\ \ref{f:MassFunc_Blue} and
\ref{f:MassFunc_Red}, respectively. The mass functions are shown in two bins
of half-mass density $\rho_{\rm h}$ in the left-hand panels, and two bins of
projected galactocentric radius $R_{\rm gal}$ in the right-hand panels. We
plot the mass functions as $dN/d\log\cM = (\cM\,\ln 10)\, dN/d\cM$ rather than
$dN/d\cM$ so as to create 
a shape similar to the familiar GC {\it luminosity\/} functions. 

\subsubsection{Mass Function of Blue GCs} 

The mass functions of the blue GCs in Fig.\ \ref{f:MassFunc_Blue} 
show the familiar single-peaked ``log-normal'' shape of
luminosity or mass functions of GCs in ``normal'' early-type
galaxies, as expected for ``ancient'' GC systems \citep[see, e.g.,][]{jord+07}. 
Furthermore, Fig.\ \ref{f:MassFunc_Blue} shows that the
peak mass exhibits a clear dependence on $\rho_{\rm h}$, whereas it
depends much less strongly on $\Rgal$. This is consistent with the situation
for cluster mass functions in our Galaxy and the Sombrero galaxy
(\citealt{chan+07}; MF08). 

To enable a systematic comparison between the different peak cluster masses
for different ranges of $\rho_{\rm h}$ and $\Rgal$, we fit the data in Fig.\
\ref{f:MassFunc_Blue} by a model 
mass function. For purposes of consistency with recent work on fitting
mass functions of ``old'' star clusters (cf.\ above), we adopt a sum of
so-called ``evolved Schechter functions'':  
\begin{equation}
\frac{dN}{d\cM} = \sum_i A_i \; \frac{1}{(\cM + \Delta_i)^2} \; \exp\left( -
  \frac{\cM + \Delta_i}{\cM_{\rm c}} \right) 
\label{eq:evsch}
\end{equation}
(\citealt{jord+07}; \citealt{chan+07}; MF08) where $\Delta_i \equiv (\mu_{\rm
  ev}\,t)_i \propto \sqrt{\rho_{\rm h,\,i}}$ is the cumulative mass loss of
cluster $i$, $\mu_{\rm ev}$ is the rate of evaporative mass loss by two-body
relaxation, $t$ is the elapsed time (taken to be 13 Gyr in this case), $A_i$
is a normalization constant for each cluster, $\cM_{\rm c}$ is the cutoff mass of
the \citet{sche76} function, and the sum is over all clusters in the
population. 
Following \citet{chan+07}, we first fitted a model with a single term
$\widetilde{\Delta}$ (the median value of $\Delta$) in equation
(\ref{eq:evsch}) to the mass functions for the full sample of 212 blue
clusters to produce a ``median'' mass function which is depicted as the dashed
curves in Fig.\  \ref{f:MassFunc_Blue}. 
After finding the best-fitting values for $\cM_{\rm c}$
and $\widetilde{\Delta}$, we determined the value of the normalization constant
$C_{\rm ev}$ in $\widetilde{\Delta} = \widetilde{\mu_{\rm ev}} \,t = 
C_{\rm ev}\, \widetilde{\rho_{\rm h}}^{1/2}$. Next, we fitted models for  
the two bins in $\rho_{\rm h}$ and the two bins in $\Rgal$ using separate
terms for each cluster in equation (\ref{eq:evsch}), using the values for
$\cM_{\rm c}$ and $C_{\rm ev}$ derived above. These models are shown by solid
curves in Fig.\ \ref{f:MassFunc_Blue}. 
The peak mass $\cM_{\rm p}$ corresponding to the turnover magnitude of GC
luminosity functions is found from the values of $\widetilde{\Delta}$ and
$\cM_{\rm c}$ by solving   
\begin{equation}
\left. \partial \left( \log \left(\frac{dN}{d\log\cM}\right) \right)
  \middle/ \partial (\log\cM) = 0 \right.
\label{eq:partial}
\end{equation}
for $\cM$, to which the solution is 
\begin{equation}
\cM_{\rm p} = \frac{-(\widetilde{\Delta}+\cM_{\rm c}) +
  \sqrt{(\widetilde{\Delta}+\cM_{\rm c})^2 +
  4 \widetilde{\Delta} \cM_{\rm c}}}{2}\mbox{.} 
\label{eq:M_p}
\end{equation}
Resulting values for $\cM_{\rm p}$, $\widetilde{\Delta}$ and $\widetilde{\rho_{\rm h}}$
for each bin in $\rho_{\rm h}$ and $\Rgal$ are listed in
Table~\ref{t:massloss}. It turns out that $\widetilde{\Delta}$ scales with
$\widetilde{\rho_{\rm h}}$ as $\widetilde{\Delta} \propto \widetilde{\rho_{\rm
    h}}^{\beta}$ with $\beta = 0.44 \pm 0.10$. This is consistent with the
predicted scaling $\Delta \propto \rho_{\rm h}^{1/2}$ for two-body relaxation,
given the exponential cutoff of the Schechter function at $\cM_{\rm c}$. 

\subsubsection{Mass Function of Red GCs} 

\begin{figure*}[tb]
\centerline{
\psfig{figure=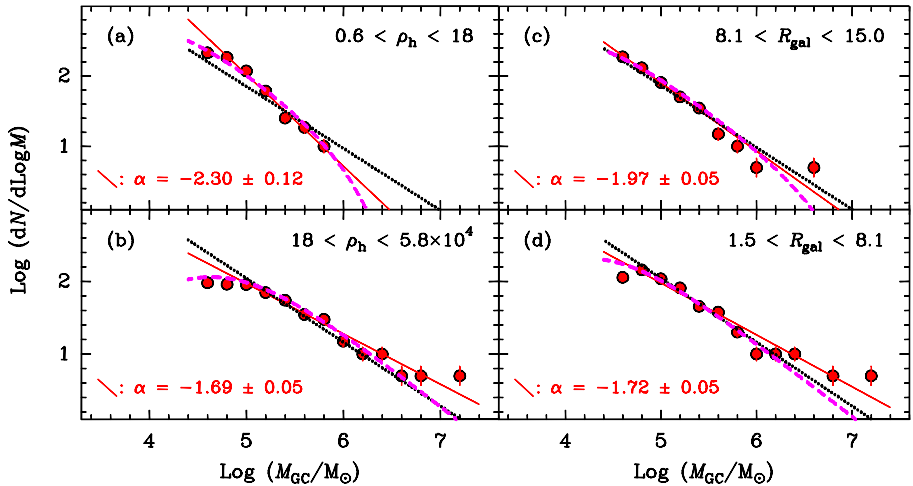,width=12cm}
}
\caption{Mass functions of red clusters in NGC~1316 with robust size
  measurements. Panels (a) and (b) show data for two ranges of half-mass
  density $\rho_{\rm h}$ in units of M$_{\odot}$ pc$^{-3}$, as indicated in
  the top right. Panels (c) and (d) show data for two ranges of projected
  galactocentric distance $\Rgal$ in kpc, as indicated in the top right. The
  (red) solid curves represent power-law fits to the mass functions for $\cM_{\rm
    GC} > 10^5$ M$_{\odot}$. The value of $\alpha$ in $dN/d\cM \propto
  \cM^{\alpha}$ is stated at the lower left of each panel. For comparison
  purposes, the (black) dotted curve in each panel represents a power-law fit
  to the mass function of the full sample of 212 red clusters, for which $\alpha =
  -1.88 \pm 0.04$. The dotted curve was normalized to the solid curve at 
  log~$(\cM_{\rm GC}/\mbox{M}_{\odot}) = 5.4$ in each panel to facilitate the
  comparison. Finally, the (magenta) dashed curves represent fits of equation 
  (\ref{eq:evsch}) to the data by summing over mass-loss terms for all
  individual clusters in the samples represented in each panel. See Sect.\
  \ref{s:massfunc}.2 for details.
\label{f:MassFunc_Red}}
\end{figure*}

Fig.\ \ref{f:MassFunc_Red} shows that the mass function of the red GCs in NGC
1316 has much in common with the power-law mass functions of young massive 
clusters. In fact, the mass function of the full sample of 212 red GCs is very
well fit by a power law $dN/d\cM \propto \cM^{\alpha}$ with $\alpha = -1.88
\pm 0.04$, consistent with those found in younger merger remnants and starburst
galaxies \citep[e.g.,][]{meur+95,fall+09}. However, splitting 
the cluster system into two bins of $\rho_{\rm h}$ and two bins of $\Rgal$
reveals more information about the dynamical state of the red GC
system. Performing power-law fits to the mass function of the four cluster
subsamples shows that the value of $\alpha$ differs systematically between the
subsamples in the sense that the slope is steepest for the lowest-density
clusters and flattest for the highest-density clusters, with the two bins in
$\Rgal$ showing intermediate values of $\alpha$. This suggests that
evaporation by two-body relaxation has already had a measurable impact on the
clusters' dynamical evolution. The apparent flattening of the mass function at
log $(\cM_{\rm cl}/\mbox{M}_{\odot}) \la 5.0$ in the high-density subsample 
(i.e., panel [b] of Fig.\ \ref{f:MassFunc_Red}) indicates this as
well. To check whether this flattening is at all consistent with the expected
shape of the cluster mass function at an age of 3 Gyr due to two-body
relaxation, we adopt the mass density-dependent mass loss rate $\mu_{\rm ev}$
found above for the full system of {\it blue\/} GCs in NGC~1316, i.e., 
\begin{equation}
\mu_{\rm ev} = 875 \, \left(\frac{\rho_{\rm h}}{\mbox{M}_{\odot}
    \,\mbox{pc}^{-3}}\right)^{1/2} \, \mbox{M}_{\odot}\, \mbox{Gyr}^{-1}
\label{eq:bluemdot}
\end{equation}
(see values of $\widetilde{\Delta}$ and $\widetilde{\rho_{\rm h}}$ for all
blue GCs in Table \ref{t:massloss}). Inserting equation (\ref{eq:bluemdot})
into equation (\ref{eq:evsch}) and assuming $t$ = 3 Gyr for the red clusters,
we fit equation (\ref{eq:evsch}) to the two bins in $\rho_{\rm h}$ and the two 
bins in $\Rgal$, using separate terms for each individual cluster as done in
the previous section. Resulting values for $\cM_{\rm p}$ for each subsample
are listed in Table~\ref{t:massloss}, and the resulting mass function fits are
shown in Fig.\ \ref{f:MassFunc_Red} as dashed curves. Note that the
fitted mass functions are consistent with the data, including the apparent
flattening of the mass function at log $(\cM_{\rm cl}/\mbox{M}_{\odot}) \la 5.0$
in the high-density subsample. 

This result is consistent with our earlier claims in \citet{goud+04}, even
though the latter study used incorrect aperture corrections for the large red
clusters, which are mainly located in the outer regions. Furthermore, we can
now quantify that the flattening of the mass function is stronger for clusters
with high mass density than those with smaller galactocentric radii, as
expected if two-body relaxation dominated the dynamical evolution (but see
Section~\ref{s:diffuse}).  

\section{Dynamical Evolution of the Intermediate-Age Cluster Population}
 \label{s:evol_iagc}

As mentioned in the Introduction, the population of red clusters
in the 3-Gyr-old merger remnant NGC~1316 provides an important
opportunity to test whether and how the power-law LFs seen in young
merger remnant galaxies may evolve into the ubiquitous
bell-shaped LFs of ancient GCs in ``normal'' galaxies. This topic is
addressed in this Section by applying dynamical evolution model
calculations from the recent literature to the observed current properties of
the red clusters in NGC~1316. To compare the results with ancient GCs in ``normal''
galaxies, we let the red clusters evolve dynamically for another 10 Gyr. Mass loss
by stellar evolution is also taken into account, using the BC03 SSP
models at solar metallicity which yield a mass loss of 9\% between ages of 3
and 13 Gyr. 

\subsection{Evaporation by Two-Body Relaxation} \label{s:evap}

We use two independent models to evaluate the effect of evaporation by
two-body relaxation to the red GCs in NGC~1316 over the next 10 Gyr. 

\begin{figure*}[tp]
\centerline{
\psfig{figure=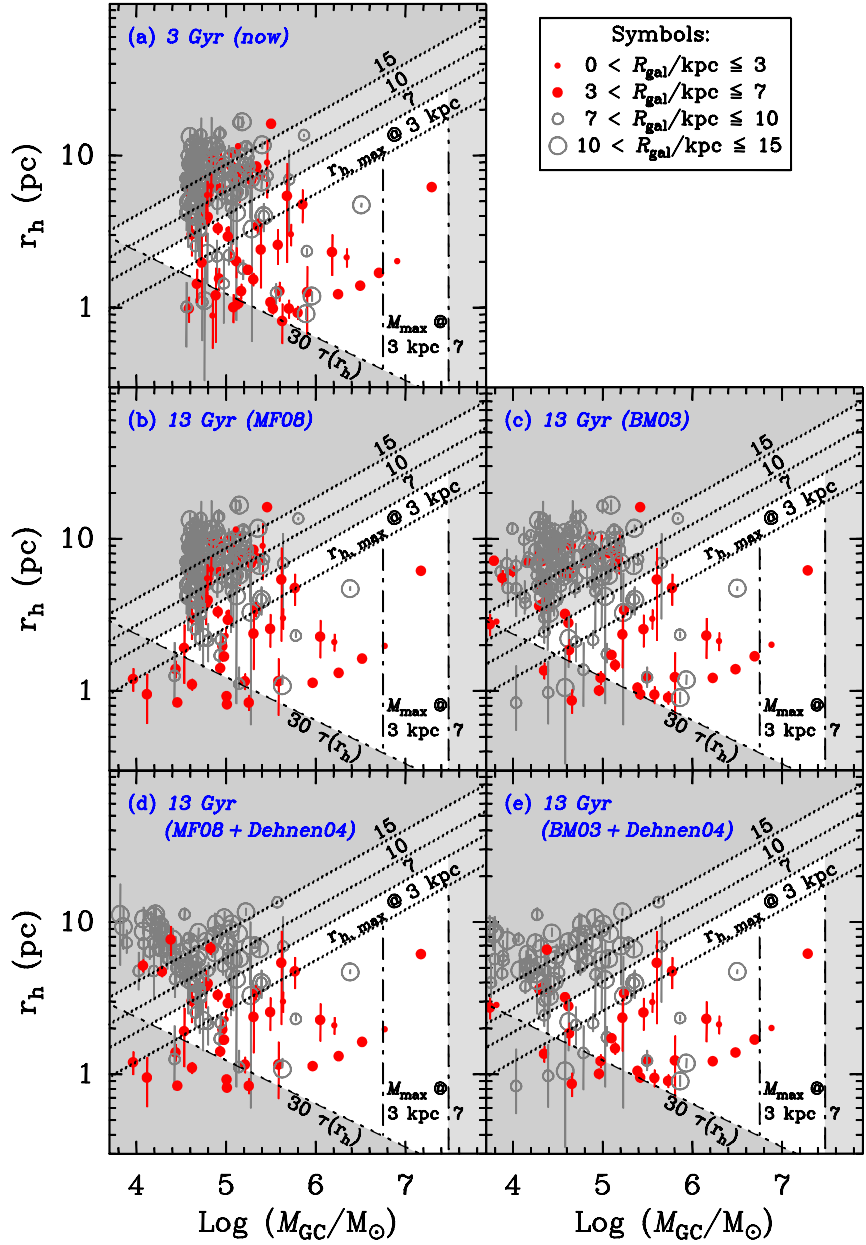,height=16.5cm}
}
\caption{Survival diagram for the red clusters in NGC~1316 as function of
  time using dynamical evolution calculations described in 
  \S\S\, \ref{s:evap} and \ref{s:diffuse}. 
  Panel (a) shows the same data as Fig.\ \ref{f:vitalplot1}a for
  reference. Panel (b) shows the surviving clusters at an age of 13 Gyr
  after applying mass loss due to two-body relaxation according to the
  prescriptions of MF08. Panel (c) is similar to panel (b), except that mass
  loss due to two-body relaxation has been applied according to the
  prescriptions of BM03. Panels (d) and (e) are similar to panels (b) and (c),
  respectively, except that mass loss due to tidal shocks according to
  prescriptions of \citet{dehn+04} has also been applied. Symbols are shown in
  the legend in the upper right. 
\label{f:vitalplot_nosegr}}
\end{figure*}

{\it (i) The McLaughlin \& Fall (2008) model}.~ 
We first use the aforementioned model of MF08 who argue
that the change in shape of the globular cluster mass function (GCMF) from
young to old systems is due mainly to evaporation by two-body
relaxation \citep[see also][]{falzha01}. Driven by their finding that the
dependence of the peak mass $\cM_{\rm p}$ of the Galactic GCMF on half-mass
density is approximately $\cM_{\rm p} \propto \rho_{\rm h}^{1/2}$, as
predicted for evaporation by internal two-body relaxation, they suggest that
evaporation by two-body relaxation is dominated by effects internal to the
clusters and ignore the influence of the tidal field of the galaxy.    
We use equation (\ref{eq:bluemdot}) derived above for the blue GCs in NGC~1316
to estimate evaporation-driven mass loss for the red GCs according to the MF08
model.
The resulting distribution of the surviving red GCs at an age of 13 Gyr in the
mass-radius plane is shown in panel (b) of Fig.\ \ref{f:vitalplot_nosegr}.

{\it (ii) The Baumgardt \& Makino (2003) model}.  
\citet[][hereafter BM03]{baumak03} performed extensive N-body simulations of
star clusters, taking a (static) Galactic tidal field into account. 
Similar to MF08, they assume that GCs fill their Roche lobe in the tidal
field. Adopting a King model with $W_0 = 7$ (i.e., $C_K \simeq 30$), we use
their equation (7) to calculate dissolution times $t_{\rm diss}$ in Gyr for
each red GC. In case $t_{\rm diss} > 10$ Gyr, remaining GC masses at an age of
13 Gyr were approximated using ${\cal{M}}_{\rm cl} \, (13 \:{\rm Gyr}) = 0.91
\, {\cal{M}}_{\rm cl,0} \, (1-10/t_{\rm diss})$ where ${\cal{M}}_{\rm cl,0}$
is the current cluster mass (see BM03). We employ circular GC orbits 
for this exercise, recognizing that dissolution times vary with orbit
eccentricity $\epsilon$ approximately as $t_{\rm diss} (\epsilon) =
(1-\epsilon)\: t_{\rm diss} (0)$ (see BM03). 
Panel (c) of Fig.\ \ref{f:vitalplot_nosegr} shows the distribution of the
surviving red GCs at an age of 13 Gyr in the mass-radius plane according to
the BM03 model. 

Comparing panels (b) and (c) of Fig.\ \ref{f:vitalplot_nosegr}, one can
see that the MF08 model yields higher mass loss for compact 
low-mass GCs while the BM03 model yields higher mass loss for larger GCs at a
given initial mass. However, it is also clear that both models predict little
mass loss for GCs with ${\cal{M}}_{\rm cl,0} \la 10^5 \; {\rm M}{_\odot}$ and $\rh \ga
7$ pc, especially the MF08 model. Note that such diffuse, low-mass clusters
are quite common in NGC~1316 even though they are rare in 
``normal'' elliptical galaxies. This is further discussed below. 

\subsection{Diffuse Red Clusters and the Influence of Tidal Shocks} 
\label{s:diffuse} 

While diffuse, low-mass, red clusters such as those seen in NGC~1316 have
hitherto not been identified in nearby ``normal'' elliptical galaxies, they
are known to exist in about a dozen nearby lenticular (S0) galaxies 
\citep[such clusters are sometimes called ``faint
fuzzies'';][]{larbro00,brolar02,peng+06b}. Studies of the 
lenticular galaxies that host such clusters commonly found various 
signatures of past interactions with neighboring galaxies \citep[][and
references therein]{brolar02,hwalee06}. 
The presence of several such clusters in the intermediate-age merger remnant
NGC~1316 seems relevant in this context, also because its body's dynamical
$v/\sigma$ ratio is consistent with an isotropic rotator \citep{arna+98},
which is a typical signature of lenticular galaxies \citep[e.g.,][]{bend88}. 

The population of such ``diffuse'' clusters in lenticular galaxies is
typically fainter than the turnover of the GC luminosity function (of
``normal'' galaxies, i.e., $M_V \simeq -7.4$, equivalent to $\Mcl \simeq
1.5\times 10^5$ M$_{\odot}$), with colors similar to the brighter ``red''
clusters \citep{brolar02,peng+06b}. The situation for diffuse clusters in
NGC~1316 is very similar, except that the mass range of the latter seem to
extend to masses higher by a factor of 2\,--\,5 (viz.\ Fig.\
\ref{f:vitalplot2}), which may in turn be due to such clusters having been
exposed to dynamical evolution processes for a significantly shorter time in
NGC~1316 than in ``normal'' galaxies. This raises the question:\ could the
diffuse red clusters seen in the merger remnant NGC~1316 represent the
precursors of their counterparts seen in ``normal'' lenticular galaxies? If
so, what would their masses be at an age of 13 Gyr?  

As discussed in the previous Section, evaporation by internal two-body relaxation
is very ineffective for diffuse clusters with such low densities. In the
absence of dynamical processes with higher mass loss  efficacy for such
clusters, the prevalence of such clusters in NGC~1316 would therefore seem at
odds with the significantly lower frequency of such clusters in ``normal''
lenticular galaxies.  
However, a key feature of such clusters in NGC~1316 is that they typically
have $\rh > r_{\rm h, max}$, i.e., their sizes are larger than the upper limit
imposed by the tidal field of the host galaxy at their $R_{\rm gal}$ (see Sect.\
\ref{s:survival} and Fig.\ \ref{f:vitalplot2}). This indicates that a
significant fraction of the stars in such clusters is located outside the
Roche lobe of the cluster, i.e., the outer limit of the ``bound'' part of such
clusters that is in equilibrium with the tidal field of the host galaxy.   
Disruption of these diffuse clusters by tidal shocks is therefore expected to
be significantly faster than for clusters that are tidally limited, i.e.,
clusters that do not extend beyond their Roche lobe. 
To our knowledge, the N-body simulation study of \citet{dehn+04} 
is the only one to date that 
addresses the dynamical evolution of this type of ``supertidal'', low-density
star cluster by galactic tides in detail. Dehnen et al.\ model 
the disruption of the diffuse Galactic cluster Palomar 5, for which $\rh
\simeq 20$ pc and $r_{\rm h} / r_{\rm h, max} \approx 2$ \citep[see
also][]{oden+03}, similar to many diffuse clusters in NGC~1316. The
simulations of \citet{dehn+04} suggest that clusters with $r_{\rm h} / r_{\rm
  h, max} > 1$ that move along eccentric orbits never become tidally
limited\footnote{This is unlike more ``typical'' surviving GCs, which
  feature higher $\rho_{\rm h}$ and experience tidal shocks on time scales that
  are long relative to their dynamical time. For such clusters, tidal shocks
  are thought to cause significant mass loss only during their first few galactic
  orbits \citep{gned+99,falzha01} or during galaxy interactions
  \citep{krui+11}. At an assumed age of 3 Gyr, the red clusters in NGC~1316
  already went through this era.}. This would mean that 
their dynamical evolution would be mainly driven by tidal shocks rather than
evaporation by two-body relaxation. 
For the case of Palomar 5 which has a current mass of $\sim 5\times 10^3$
M$_{\odot}$, \citet{dehn+04} find that a cluster with an orbit-averaged mass
loss $\dot{\cal{M}}_{\rm cl} = 5\times 10^3$ M$_{\odot}$ Gyr$^{-1}$ yields the best fit. For an
assumed age of 13 Gyr for Palomar 5, this implies a mass of $5.5\times 10^4$
M$_{\odot}$ at an age of 3 Gyr assuming a constant Galactic tidal field
strength. Note that this mass is just above the low-mass end of the diffuse
clusters in NGC~1316 (with reliable size measurements). 
Using the values of $\dot{\cal{M}}_{\rm cl}/{\cal{M}}_{\rm cl,\,0}$ as
function of $r_{\rm h} / r_{\rm h, max}$ and $W_0$ from the simulations of
\citet{dehn+04} (using polynomial interpolation), we estimate
masses at an age of 13 Gyr for the clusters in NGC~1316 with $r_{\rm h} /
r_{\rm h, max} > 1$ for the case of 
$W_0 = 4.2$. 
To do so, the values of $\dot{\cal{M}}_{\rm cl}/{\cal{M}}_{\rm cl,\,0}$ from
Dehnen et al.\ were scaled for a given cluster by considering that the mass 
loss timescale due to tidal shocks $\tau_{\rm sh}$ scales as
$\tau_{\rm sh} \propto P_{\rm cl}$ \citep[e.g.,][]{gned+99}. Here $P_{\rm cl}$
is the period of the cluster's orbit, estimated as $P_{\rm cl} \approx 2\pi
\Rgal^{3/2} (G\cM_{<R_{\rm gal}})^{-1/2} \approx 2 \pi \Rgal (V_{\rm
  circ}^2/3)^{-1/2}$, where $\cM_{<R_{\rm gal}}$ is the mass of NGC~1316
inside the galactocentric distance of the cluster in question. In this context
we assume $\Rgal = 16.4$ kpc for Palomar 5, which is the equivalent
radius of its eccentric orbit \citep{oden+03}. The results are
shown in panels (d) and (e) of Fig.\ \ref{f:vitalplot_nosegr}.  

Note that the application of the \citet{dehn+04} calculations indeed predicts
that many of the diffuse red clusters in NGC~1316 will end up in the area
within the survival diagram known to be occupied by diffuse low-mass clusters
in our Galaxy (several `Palomar'-type clusters; see panel (c) of Fig.\
\ref{f:vitalplot1}). Hence, this yields a low-mass tail to the cluster mass
function that is lacking when only applying cluster disruption due to two-body
relaxation (especially for the MF08 model).  

\begin{figure}[tb]
\centerline{
\psfig{figure=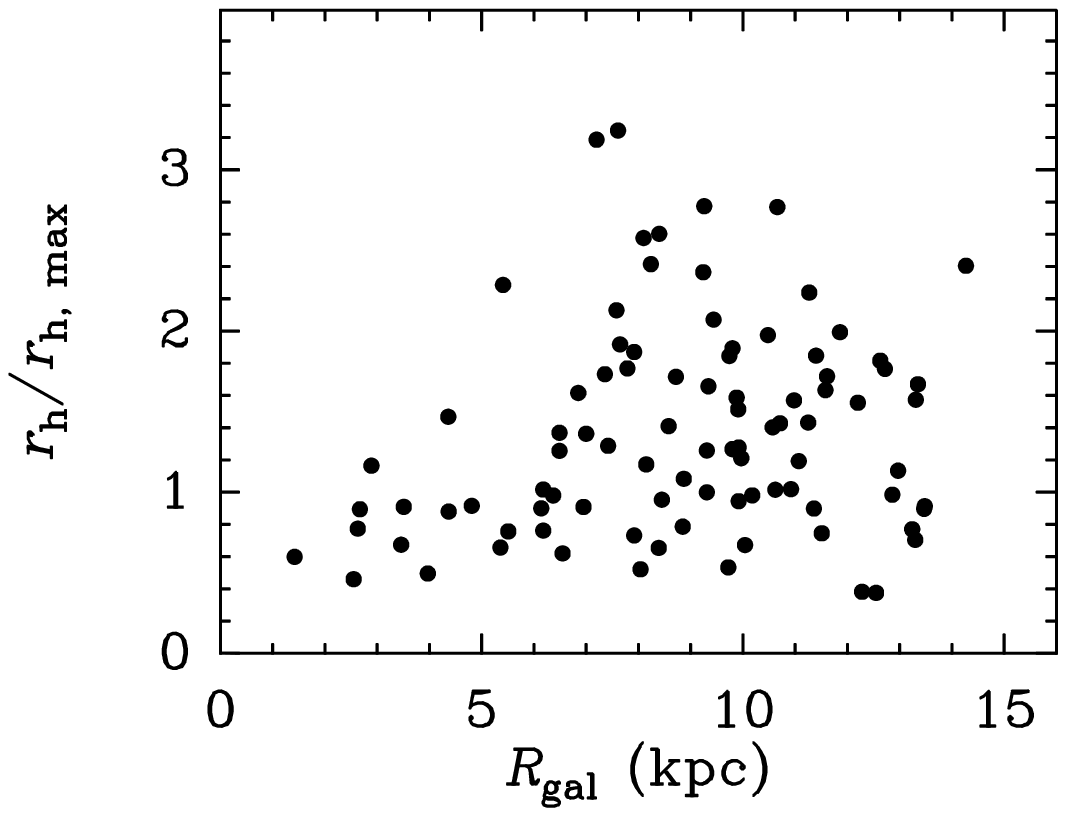,width=7cm}
}
\caption{$\rh/r_{\rm h,\,max}$ as function of galactocentric distance for red
  clusters with $4.6 < \log ({\cal{M}}_{\rm cl}/M_{\odot}) < 4.8$. Note the
  scarcity of ``supertidal'' clusters with $\rh/r_{\rm h,\,max} \ga 2$ at
  $\Rgal \la 5$ kpc. 
\label{f:Rgal_vs_rh_rhmax}}
\end{figure}

\begin{figure}[tb]
\centerline{
\psfig{figure=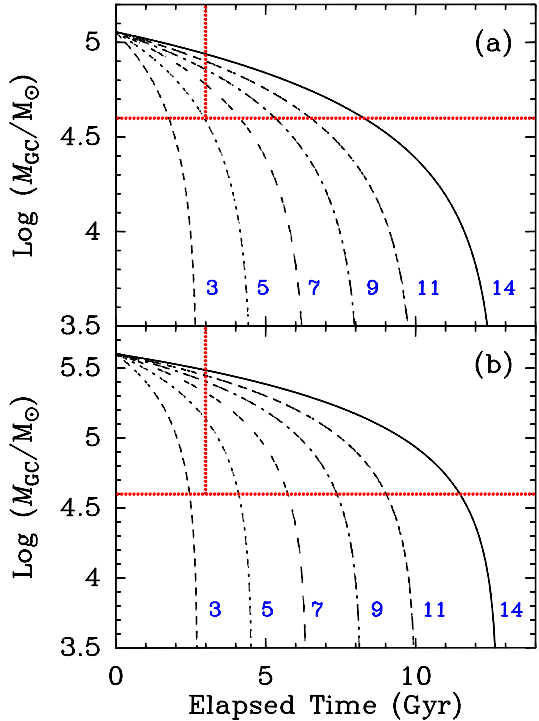,width=7cm}
}
\caption{{\it Panel (a):\/} Dynamical evolution due to tidal shocking according to
  simulations of \citet{dehn+04} for a cluster with $\rh/r_{\rm h,\,max} =
  2$ and   $\log({\cal{M}}_{\rm cl,\,0}/M_{\odot}) = 5.06$ for six values of
  galactocentric distance $\Rgal$ in NGC~1316. The values of $\Rgal$ in kpc
  are indicated near the bottom of the plot, to the right of their respective
  associated curves.  
  The horizontal dotted line indicates the low-mass cutoff of the nominal
  cluster sample. The vertical dotted line indicates an elapsed time of 3 Gyr,
  the assumed age of the red clusters in NGC~1316. {\it Panel (b):\/} Similar to
  panel (a), but for $\log({\cal{M}}_{\rm cl,\,0}/M_{\odot}) = 5.60$. 
  See discussion in Sect.\ \ref{s:diffuse}. 
\label{f:detectionplot}}
\end{figure}

An interesting side consequence of the results of the \citet{dehn+04} simulations
is that they would explain why the supertidal clusters with masses close to
the low-mass cutoff implied by the constraint of avoiding biases related to
varying incompleteness (i.e., $V < 25.3$ mag) are only found in the outer
regions of NGC~1316. To illustrate this, we plot $\Rgal$ versus the ratio
$\rh/r_{\rm h,\,max}$ in Figure~\ref{f:Rgal_vs_rh_rhmax} for all red clusters
with $25.3 > V > 24.8$ mag, corresponding to $4.6 < \log ({\cal{M}}_{\rm
  cl}/M_{\odot}) < 4.8$. This plot shows clearly that diffuse clusters with
$\rh/r_{\rm h,\,max} \ga 2$ are only found at $\Rgal \ga 5$ kpc. Taking the
median value of $\Rgal$ for clusters with $4.6 < \log ({\cal{M}}_{\rm
  cl}/M_{\odot}) < 4.8$ and $1.8 < \rh/r_{\rm h,\,max} < 2.2$ as a proxy
($\widetilde{\Rgal}$ = 9.0 kpc), we obtain a typical value of
$\dot{\cal{M}}_{\rm cl}/{\cal{M}}_{\rm cl,\,0}$ = $-$0.12~Gyr$^{-1}$ from the
Dehnen et al.\ simulations for the case of $W_0 = 4.2$ and $\rh/r_{\rm
  h,\,max} = 2.0$. For a 3-Gyr-old cluster with $\log({\cal{M}}_{\rm
  cl}/M_{\odot}) = 4.80$, we then obtain $\log({\cal{M}}_{\rm
  cl,\,0}/M_{\odot}) = 5.06$ at an age of $\approx 10^8$ yr (i.e., after the
bulk of mass loss due to stellar evolution has already
occurred). Figure~\ref{f:detectionplot}a shows ${\cal{M}}_{\rm cl}$ versus 
time for such a cluster for six values of $\Rgal$ 
assuming $\dot{\cal{M}}_{\rm cl}/{\cal{M}}_{\rm cl,\,0}$ = $-$0.12~Gyr$^{-1}$
at $\Rgal = 9$ kpc. The low-mass cutoff of $\log({\cal{M}}_{\rm cl}/M_{\odot})
= 4.6$ is indicated by a dotted horizontal line. Note that the Dehnen et al.\
simulations predict that such a cluster would only remain in a
magnitude-limited sample with $V < 25.3$ mag at an age of 3 Gyr for $\Rgal \ga
5$ kpc, consistent with what we see. This suggests that {\it the absence of
  such low-mass ``supertidal'' clusters in the inner regions of NGC~1316 is
  caused by tidal shocking\/} rather than due to (e.g.) completeness-related
issues. Incidently, this could also be the reason why \citet{larbro00} find a
deficit of ``faint fuzzies'' in the central regions of the lenticular galaxy
NGC~1023. For comparison purposes, Figure~\ref{f:detectionplot}b shows the
same as Figure~\ref{f:detectionplot}a but for $\log({\cal{M}}_{\rm
  cl,\,0}/M_{\odot}) = 5.6$, which yields a cluster with a mass at age = 3 Gyr
that is similar to the maximum mass attained by diffuse red clusters in
NGC~1316. Note that such clusters are predicted to stay detectable in the
outer regions of NGC~1316 for 10-12 Gyr, which is consistent with them having
been detected in the outskirts of some nearby ``normal'' S0 galaxies. 

We note that the simulations of \citet{dehn+04} only covered a limited
parameter space in ${\cal{M}}_{\rm cl}$, $r_{\rm h} / r_{\rm h, max}$ and
$W_0$, and they used a gravitational potential model for our Galaxy. Hence, these
results should be used with caution in a quantitative sense. 
Unfortunately, other N-body simulations of tidal stripping of clusters
that extend beyond their Roche lobe are still lacking in the literature to our
knowledge. 
Given the results shown here, it seems quite useful to perform such
simulations for clusters with a range of properties encompassing those found
for the diffuse red clusters in NGC~1316 as well as a more appropriate
galactic potential (e.g., using a more massive bulge component). Such 
calculations would yield relevant (more quantitative) insights into the
important question whether these clusters may indeed evolve into the
``extended clusters'' or ``faint fuzzies'' found in 
older lenticular  galaxies which often show lingering evidence of interactions
with other galaxies (e.g., nearby neighbors, see 
\citealt{brolar02,peng+06b}). The nature of diffuse red clusters is further
discussed in Sect.\ \ref{s:disc_DRCs}. 

\subsection{Resulting Mass Functions of Red GCs} \label{s:evol_MFs}

In the context of building mass functions for the red clusters at an age of 13
Gyr from the dynamical evolution calculations mentioned above, we recall that
so far, we only considered clusters with $V \leq 25.3$ mag in order to avoid
biases related to varying incompleteness for fainter clusters (see
Section \ref{s:rh_vs_Rgal}). However, it is likely that the assumed magnitude
limit has an impact on the resulting mass function at low masses. To
evaluate this impact, we build mass functions for two magnitude limits:
$V \leq 25.3$ mag (as before) and $V \leq 25.8$ mag. For reference, the
completeness fraction in the latter (fainter) cluster sample is $\ga 0.20$ as
opposed to $\ga 0.46$, and the minimum S/N returned by the {\sc ishape} fits
is $\sim$\,22 as opposed to $\sim$\,50.  

\begin{figure*}[tb]
\centerline{
\psfig{figure=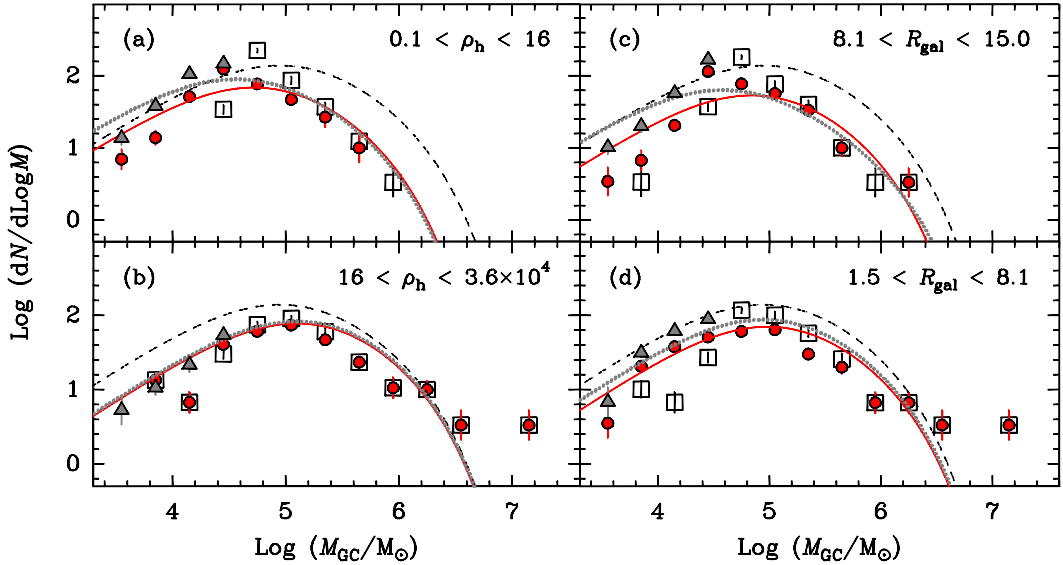,width=14cm}
}
\caption{Predicted mass functions of red clusters in NGC~1316 shown in panel
  (c) of Fig.\ \ref{f:vitalplot_nosegr} at an age of 13 Gyr,  
  using MF08 prescriptions for mass loss by two-body relaxation. Panels (a)
  and (b) show data for two ranges of half-mass density $\rho_{\rm h}$ in
  units of M$_{\odot}$ pc$^{-3}$, as indicated in the legend. Panels (c) and
  (d) show data for two ranges of projected galactocentric distance $\Rgal$ in
  kpc as indicated in the legend. Open squares represent results using {\it
    only\/} mass loss due to two-body relaxation, while filled (red) circles
  represent results that take into account additional mass loss due to tidal
  shocking according to simulations of \citet{dehn+04}. The solid (red) curves
  represent fits of equation (\ref{eq:evsch}) to the filled circles by
  summing over mass-loss terms for all individual clusters within the bins of
  $\rho_{\rm h}$ or $\Rgal$ mentioned in each panel. 
  For comparison purposes, the dashed (black) curve in each panel was computed
  from equation (\ref{eq:evsch}) with a single term using the median value of
  $\rho_{\rm h}$ for the full sample of 192 surviving red clusters. Finally,
  the filled (grey) triangles and the dotted lines depict the same results as
  the filled circles and the solid lines, respectively, but now for red
  clusters with $V < 25.8$ instead of $V < 25.3$. See discussion in Section
  \ref{s:evol_MFs}.
\label{f:MassFunc_MF08}}
\end{figure*}

\begin{figure*}[tb]
\centerline{
\psfig{figure=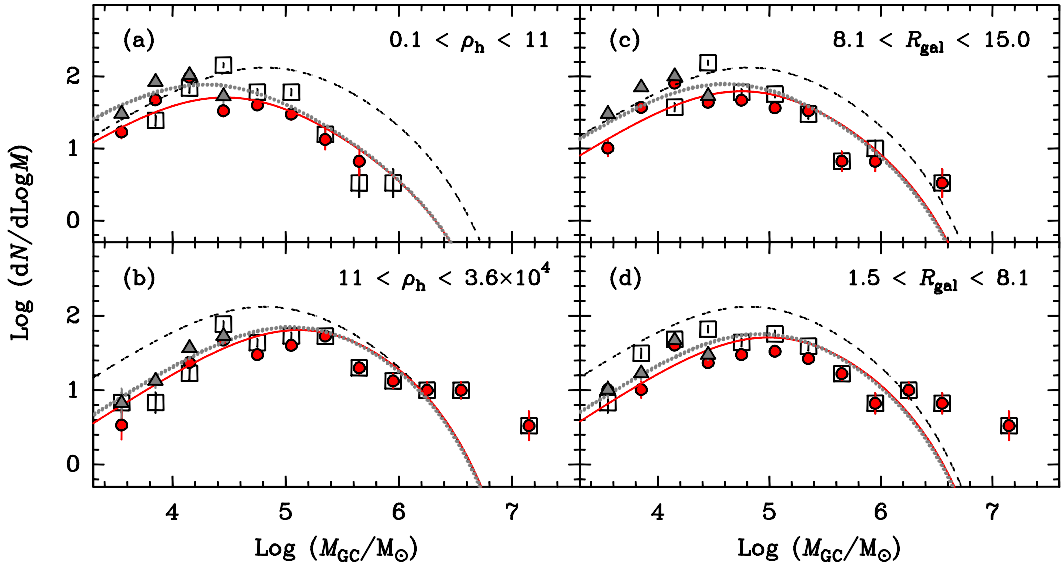,width=14cm}
}
\caption{Same as Fig.\ \ref{f:MassFunc_MF08}, but now for the red clusters
  shown in panels (c) and (e) of Fig.\ \ref{f:vitalplot_nosegr}, i.e., using
  the BM03 prescriptions for mass loss by two-body relaxation. 
\label{f:MassFunc_BM03}}
\end{figure*}

The completeness-corrected cluster mass functions resulting from the
calculations mentioned above are shown in Figs.\ \ref{f:MassFunc_MF08} and
\ref{f:MassFunc_BM03} (using two-body relaxation prescriptions from MF08 and
BM03, respectively). Mass functions are again shown for two bins of  
$\rho_{\rm h}$ in the left-hand panels, and two bins of 
$R_{\rm gal}$ in the right-hand panels. 
The curves drawn in Figs.\ \ref{f:MassFunc_MF08} and \ref{f:MassFunc_BM03}
represent fits of equation (\ref{eq:evsch}) to the cluster masses evolved to
an age of 13 Gyr using cluster disruption due to both two-body relaxation and tidal shocks 
(i.e., the filled symbols), performed in a manner identical to that described
in Section~\ref{s:massfunc} and shown in Fig.\ \ref{f:MassFunc_Blue}. 
Resulting values for $\cM_{\rm p}$, $\Delta$ and $\widetilde{\rho_{\rm h}}$
for each bin in $\rho_{\rm h}$ and $\Rgal$ are again listed in 
Table~\ref{t:massloss}.  

Figs.\ \ref{f:MassFunc_MF08} and \ref{f:MassFunc_BM03} show several items of
interest. First of all, the estimated mass functions for an age of 13 Gyr  
appear roughly similar to the familiar shape of luminosity or mass functions of
GCs in ``normal'' early-type galaxies.
Secondly, Figs.\ \ref{f:MassFunc_MF08} and \ref{f:MassFunc_BM03} both show
that $\cM_{\rm p}$ exhibits a clear dependence on $\rho_{\rm h}$, whereas it
depends less strongly on $\Rgal$. This statement is reinforced by the fact
that the dependence on $\rho_{\rm h}$ is stronger for the sample with $V \leq
25.8$ mag than for that with $V \leq 25.3$ mag. Quantitatively, the mass
function fits to the red clusters with $V \leq 25.8$ mag imply that
$\widetilde{\Delta} \propto \widetilde{\rho_{\rm h}}^{\beta}$ with $\beta =
0.41\pm0.10$ when applying the MF08 prescriptions for two-body relaxation, and
$\beta = 0.44\pm0.10$ when applying the BM03 prescriptions. This is again
consistent with the situation for cluster mass functions in our Galaxy and the
Sombrero galaxy (\citealt{chan+07}; MF08) as well as with our findings for the
``blue'' metal-poor GCs in NGC~1316 (see Sect.\ \ref{s:massfunc} and Fig.\
\ref{f:MassFunc_Blue}), and hence with the notion that dynamical evolution by 
two-body relaxation is an important mechanism in shaping cluster mass
functions after the early era of rapid mass loss by stellar evolution and gas
expulsion (e.g., MF08).  However, the shapes of the estimated mass functions
are significantly better matched to those of the overplotted model fits
(evolved Schechter functions) for the calculations that take into account
disruption of the diffuse ``supertidal'' clusters due to tidal shocks than
those that do not. This effect is illustrated most clearly in Fig.\
\ref{f:MassFunc_MF08}a where the calculations that use the MF08 model of mass
loss by two-body relaxation (i.e., the open squares) are poorly fit by
evolved Schechter functions (Eq.\ \ref{eq:evsch}) at masses $\la 10^5$
\Msun. This suggests that the observed functional dependence of $\cM_{\rm 
  p}$ on $\rho_{\rm h}$ among ancient cluster systems is {\it not necessarily
  due to evaporation by two-body relaxation alone}. Our analysis
suggests that it may be due in part to disruption of very low-density
``supertidal'' clusters by tidal shocks. 
The reason why the effect of disruption of supertidal  clusters by tidal
shocks can ``masquerade'' as  that of two-body relaxation in mass function
plots is that both mechanisms are  characterized by mass loss rates that are
highly linear with time (see \citealt{dehn+04,jord+07}).  

Finally, we comment on the impact of the possibility that a small
fraction of the fainter red GCs in NGC~1316 are in fact ``old''
($\simeq$ 13 Gyr rather than 3 Gyr old; cf.\ Section~\ref{s:data}) on
the mass functions shown in Figures~\ref{f:MassFunc_MF08} and
\ref{f:MassFunc_BM03}. Since ancient red GCs in ``normal'' early-type
galaxies typically have radii of 1\,--\,4 pc
\citep[e.g.,][]{jord+05,paol+11}, Figure~\ref{f:vitalplot_nosegr}
shows that the impact of a small fraction of red GCs being old is
predicted to be insignificant. Specifically, the dynamical evolution
from 3 Gyr to 10 Gyr performed in this Section would have moved of
order 7\,--\,10 of the red GCs with $1 \la \rh \la 4$ pc incorrectly
to lower mass bins, but the mass function at those lower masses is
entirely dominated by the GCs with $\rh \ga 5$ pc since the tidal
shocks in the \citet{dehn+04} model are only effective for the latter
GCs.

\section{Summary and Discussion}  \label{s:disc}

Using {\it HST/ACS\/} images, we have conducted size measurements of GCs in the
giant $\sim$\,3-Gyr-old merger remnant galaxy NGC~1316 to obtain insights on
the dynamical status of its ``blue'' (metal-poor) and ``red'' (metal-rich) GC
subpopulations. Using simulations and calculations of cluster disruption from
the recent literature, we make predictions on the effect of 10 Gyr of
dynamical evolution of the red GCs which are assumed to have an age of 3 Gyr
based on earlier work. These results are used in the context of evaluating the
scenario that the ``red'' GCs in ``normal'' giant early-type galaxies (along
with their bulge component) were formed in a way similar to that observed in
gas-rich galaxy mergers today.  

\subsection{Dynamical Status of the Blue GCs} \label{s:disc_bGCs}

For the blue GCs, we find that their properties are consistent with those
typically found for blue GCs in ``normal'' giant early-type galaxies. The GC
mass function (MF) is consistent with an lognormal  distribution, similar to
the MF of ancient GCs in normal giant galaxies. The peak mass $\cM_{\rm p}$ of
the MF of the blue GCs increases with half-mass density 
$\rho_{\rm h}$ as  $\cM_{\rm p} \propto \rho_{\rm h}^{0.44\pm0.10}$ whereas it
stays approximately constant with $R_{\rm gal}$.   
As found recently for ancient GCs in the Milky Way and the Sombrero galaxy
\citep{chan+07,mclfal08}, the mass functions of the blue GCs are consistent
with a simple scenario in which the clusters formed with a Schechter initial
mass function and evolved subsequently by disruption driven mainly by internal
two-body relaxation (but see the last paragraph of Section~\ref{s:disc_rGCs}
below).    

\subsection{Dynamical Status of the Red GCs} \label{s:disc_rGCs}

For the intermediate-age metal-rich (red) GCs, we find 
that the faint end of the previously reported luminosity function of the
clusters outside a galactocentric radius of 9 kpc \citep[which showed a
power-law,][]{goud+04} is due to those low-mass clusters having half-light
radii large enough for disruption by two-body relaxation to be
ineffective. This is especially the case when considering the MF08
prescriptions for mass loss due to two-body relaxation. 
Moreover, we find that the radii of many of these diffuse red GCs are
larger than the theoretical maximum value imposed by the tidal field of
NGC~1316 at their $\Rgal$. We therefore suggest that the diffuse
``supertidal'' clusters in NGC~1316 lose a large fraction of their mass due to
tidal shocks in a way similar to that described by the simulations of
\citet{dehn+04} which were performed to understand the properties of the
extended cluster Palomar 5 in our Galaxy. Application and scaling of the
Dehnen et al.\ simulations to these ``supertidal'' clusters in NGC~1316 in addition
to the application of cluster disruption by two-body relaxation yields several
findings relevant to the nature of the red GCs:  
\begin{enumerate}
\item 
Clusters larger than twice the maximum radius imposed by the tidal field of
NGC~1316 with $\log (\cM_{\rm cl}/M_{\odot}) \la 5.1$ at
an age of $\approx$\,10$^8$ yr (i.e., after the era of strong mass loss due to
stellar evolution) would not survive to an age of 3 Gyr in NGC~1316 if located
within $\Rgal \la 5$ kpc, which is consistent with the observations. Note that
such clusters with $\log (\cM_{\rm cl}/M_{\odot}) \ga 4.6$ {\it would\/} be
detectable there according to our completeness calculations. 
\item
Another 10 Gyr of simulated 
dynamical evolution of the red GCs yields a MF whose shape is
similar to the lognormal shape of MFs of ancient GCs in normal galaxies
{\it throughout the range of galactocentric distances covered by the
observations}. The peak mass $\cM_{\rm p}$ of the simulated MF of the red GCs
at an age of 13 Gyr increases with $\rho_{\rm h}$ as $\cM_{\rm p} \propto
\rho_{\rm h}^{0.41\pm0.10}$ or $\cM_{\rm p} \propto \rho_{\rm
  h}^{0.44\pm0.10}$ when using the prescriptions for mass loss due to two-body
relaxation by MF08 or BM03, respectively. While this dependence of $\cM_{\rm
  p}$ on $\rho_{\rm h}$ is consistent with that found for the (old) blue GCs
in NGC~1316 as well as for the GC systems of the Milky Way and the Sombrero
galaxy as mentioned above, we emphasize that it is {\it not only due to
  internal two-body relaxation for the red GCs in NGC~1316}. With this in
mind, we suggest that disruption of low-density ``supertidal'' GCs by tidal
shocks may be partly responsible for this scaling relation among MFs of
ancient GC systems.  
\end{enumerate}

\subsection{The Nature of Diffuse Red Clusters} \label{s:disc_DRCs}

Finally, we comment on the nature and demography of the class of diffuse red  
clusters, which are quite prevalent in NGC~1316. Diffuse clusters with 
properties (luminosities, colors, sizes) similar to those found here in
NGC~1316 were first introduced as ``faint fuzzies'' by \citet{larbro00} and
are also sometimes referred to as ``extended clusters'' \citep{huxo+08}. While
extended clusters are also found in halos of nearby spiral galaxies and in 
dwarf galaxies, the great majority of the latter are metal-poor
\citep[][and references therein]{chan+04,huxo+08,geor+09a}, whereas diffuse
clusters found in NGC~1316 and younger merger remnants are metal-rich. We
therefore constrain the following discussion to the case of {\it metal-rich\/}
diffuse GCs, i.e., diffuse GCs with $(V-I)_0 \ga 1.05$ or $(g-z)_0 \ga 1.10$,
similar to the ``faint fuzzies'' originally introduced by \citet{larbro00} and
the ``diffuse star clusters'' discussed by \citet{peng+06b}. Here we
adopt the acronym DRC (``diffuse red clusters'') to describe this class
of objects. 

As shown by \citet{lars+01} and \citet{peng+06b}, ``normal'' giant early-type
galaxies that are known to host significant numbers of DRCs share a number of
properties that seem relevant to the nature of DRCs: {\it (i)\/} Their
morphological type is virtually always S0 rather than E; {\it (ii)\/} They
exhibit various signatures of ongoing or past interactions with neighboring
galaxies such as strong fine structure, disturbed morphology, the presence of
companion spiral galaxies, or their location as the dominant galaxy of a small
galaxy group. In contrast, giant S0 galaxies in the 
samples of \citet{lars+01} and \citet{peng+06b} that do not show such
signatures of past interactions do not possess any significant number of
DRCs. This observation is consistent with the suggestion by
\citet[][and references therein]{brun+09,brun+11} that such
diffuse clusters may be formed by merging of star cluster complexes such as
those commonly found in starburst regions within galaxy mergers or in the
disks of giant spiral galaxies \citep[e.g.][]{bast+05}. The fact that NGC~1316 
possesses a large number of DRCs also corroborates this idea, given its nature
as an obvious merger remnant and its dynamical nature as a rotationally
supported galaxy (similar to S0 galaxies). 

Several properties of DRCs are consistent with their denser counterparts
(i.e., red GCs of ``normal'' size, $1 \la \rh/\mbox{pc} \la 7$), which may
help further our understanding of their nature. 
The spatial distribution of DRCs in S0 galaxies typically follows that of the
galaxy light: It is very elongated for ``late-type'' S0s such as NGC~1023 and
NGC~3384 \citep{larbro00}, and more spherical for bulge-dominated S0s such as
M85 \citep{peng+06b} as well as NGC~1316 (see Sect.\ \ref{s:diffuse}). Apart
from their apparent absence in the innermost regions of galaxies, the DRCs
seem to be similar in this respect to ``normal'' red GCs, which are typically 
closely associated with the bulge component of their host galaxy as well 
\citep[e.g.,][]{kiss+97,goud+01a,goud+07,peng+06a}. In the scenario where the
red GCs and the bulges of early-type galaxies are formed during dissipative
galaxy mergers, this physical association indicates that the progenitors of
the red GCs underwent the same violent relaxation as did the stars of the
progenitor galaxies.   

The main difference between DRCs and ``normal'' red GCs within a given
normal galaxy (apart from their sizes) is that DRCs typically have masses
lower than the mass associated with the typical turnover magnitude of the GCLF
of ``normal'' red GCs \citep{brolar02,peng+06b}. This is the case in NGC~1316 as
well, although the DRCs in NGC~1316 reach slightly higher masses, perhaps due
to their younger age which renders their masses to be closer to their initial
masses. 
This difference in mass ranges is relevant because the formation efficiency of
young massive clusters in a certain region has been shown to scale with the
star formation rate (SFR) in that region \citep{larric00}. This correlation is
likely a consequence of the correlations between SFR and gas density
\citep{kenn98} and between gas density and the star formation efficiency in
high-mass clouds \citep{efre94,elmefr97}. Hence, the number of GCs of a
certain class above a certain mass threshold should be proportional to the SFR
at the time those GCs were formed 
\citep[see also][]{lars02,weid+04,bast08,geor+12}. Since it is hard to
preferentially destroy massive GCs while preserving low-mass ones, we
suggest that the typically lower masses of DRCs with respect to ``normal'' red
GCs reflects that the DRCs were formed in {\it sites with lower SFR\/} than
in the formation sites of the ``normal'' GCs. Perhaps DRCs are formed
in gas clouds with relatively low density with respect to the clouds that
produce normal GCs. This is consistent with the suggestion by \citet{elme08}
that DRCs are formed in clouds that are at or near the low-density limit for
cluster boundedness in regions with high turbulence (Mach numbers). Such
circumstances can likely occur in galaxy mergers by means of shocked clouds or
direct ISM impacts between components of the progenitor galaxies, perhaps
mainly outside the inner, dynamically most violent region of the interaction.  

Incidently, such a scenario of environment-driven formation of DRCs may also
explain why DRCs seem to be largely absent in giant E galaxies (as opposed to
S0 galaxies). According to the extensive N-body/hydro simulations of
\citet{mihher96}, mergers of two early-type disk galaxies (i.e., galaxies with
significant bulge components) typically experience peak SFRs that are factors
2\,--\,5 higher than do mergers of two late-type disk galaxies (without
bulges), with the value of the factor depending on the geometry of the galaxy
encounters. Perhaps the assembly histories of giant E galaxies with their
pressure-supported stellar dynamics were dominated by interactions involving
progenitor galaxies with pre-existing bulge components, while
rotation-supported giant S0 galaxies may have been assembled mainly (or more
predominantly) by means of interactions involving bulgeless disk galaxies. The
lower gas densities achieved in the latter interactions may have triggered the
formation of several DRCs whereas the high-SFR interactions between galaxies
with pre-existing bulges may not have. 

In the context of this scenario, the absence of DRCs in 
many giant S0 galaxies may be due to gradual disruption of DRCs by tidal
shocks as discussed above for NGC~1316. The DRCs seen in some S0 galaxies may
have been formed during relatively recent dissipative interactions of galaxies
that did not host significant bulges. Looking at
Figure~\ref{f:detectionplot}b, ``relatively recent'' in this context would
mean less than roughly 7\,--\,9 Gyr ago for a galaxy with a mass similar to
NGC~1316, and possibly longer ago for less massive 
galaxies.  For most giant early-type galaxies however, such interactions have
likely happened long enough ago for the DRCs to have suffered enough mass loss
to be rendered undetectable (or too faint to obtain reliable size information)
at the present day, at least in their inner regions which were typically
targeted by the {\it HST\/} observations. Time will tell whether DRCs can
(still) be detected and identified in the tidally benign outer regions of
giant early-type galaxies.

\paragraph*{Acknowledgments.}~The author thanks Fran\c{c}ois Schweizer for
kindly providing useful feedback on an early version of this paper. PG also 
acknowledges useful discussions with Rupali Chandar, Iskren Georgiev, Mark
Gieles, and Thomas Puzia. 
The referee is acknowledged for insightful 
comments that improved the paper.
Support for {\it HST\/} programs GO-9409 and GO-11691 was provided by
NASA through grants from the Space Telescope Science Institute, which
is operated by the Association of Universities for Research in
Astronomy, Inc., under NASA contract NAS5--26555.  STSDAS and PyRAF
are products of the Space Telescope Science Institute, which is
operated by AURA for NASA.  The SAO/NASA Astrophysics Data System was
used heavily while this paper was written. 

{\it Facility:} \facility{HST (ACS)}

\begin{table*}[tbh]
\begin{center}
\footnotesize
\caption{Properties of Evolved Schechter Function Fits to Cluster Samples.}
 \label{t:massloss}
\begin{tabular}{@{}crrcc@{}}
\multicolumn{3}{c}{~} \\ [-2.5ex]   
 \tableline \tableline
\multicolumn{3}{c}{~} \\ [-1.8ex] 
\multicolumn{1}{c}{Sample} & \multicolumn{1}{c}{Subsample} & 
 \multicolumn{1}{c}{$\widetilde{\rho_{\rm h}}$} & $\widetilde{\Delta}$ & $\cM_{\rm p}$  \\ 
\multicolumn{1}{c}{(1)} & \multicolumn{1}{c}{(2)} & \multicolumn{1}{c}{(3)}  & (4) & 
 (5) \\ [0.5ex] \tableline  
\multicolumn{3}{c}{~} \\ [-2.ex]              
Blue, $V < 25.3$ &                    all &  164 & $1.0\times 10^5$ & $8.4\times 10^4$ \\
                 &  low-$\rho_{\rm h}$ bin &   36 & $5.5\times 10^4$ & $5.0\times 10^4$ \\
                 & high-$\rho_{\rm h}$ bin & 1317 & $2.7\times 10^5$ & $1.9\times 10^5$ \\
                 &      inner $\Rgal$ bin &  229 & $1.1\times 10^5$ & $9.1\times 10^4$ \\
                 &      outer $\Rgal$ bin &  120 & $7.0\times 10^4$ & $6.2\times 10^4$ \\ [1ex]
%
Red (now), $V < 25.3$ &  low-$\rho_{\rm h}$ bin &   6 & $6.2\times 10^3$ & $6.2\times 10^3$ \\
                      & high-$\rho_{\rm h}$ bin & 274 & $4.4\times 10^4$ & $4.4\times 10^4$ \\
                      &      inner $\Rgal$ bin &  42 & $1.7\times 10^4$ & $1.7\times 10^4$ \\
                      &      outer $\Rgal$ bin &  10 & $8.4\times 10^3$ & $8.4\times 10^3$ \\ [1ex]
%
Red (13 Gyr, MF08), $V < 25.3$ &                    all &  16 & $1.0\times 10^5$ & $9.1\times 10^4$ \\
                               &  low-$\rho_{\rm h}$ bin &   4 & $6.0\times 10^4$ & $5.7\times 10^4$ \\
                               & high-$\rho_{\rm h}$ bin & 213 & $1.5\times 10^5$ & $1.3\times 10^5$ \\
                               &      inner $\Rgal$ bin & 115 & $1.1\times 10^5$ & $1.0\times 10^5$ \\
                               &      outer $\Rgal$ bin &   9 & $8.3\times 10^4$ & $7.7\times 10^4$ \\ [1ex]
%
Red (13 Gyr, BM03), $V < 25.3$ &                    all &  11 & $8.0\times 10^4$ & $7.3\times 10^4$ \\
                               &  low-$\rho_{\rm h}$ bin &   3 & $3.2\times 10^4$ & $3.1\times 10^4$ \\
                               & high-$\rho_{\rm h}$ bin & 283 & $1.5\times 10^5$ & $1.3\times 10^5$ \\
                               &      inner $\Rgal$ bin &  97 & $1.1\times 10^5$ & $1.0\times 10^5$ \\
                               &      outer $\Rgal$ bin &   6 & $6.1\times 10^4$ & $5.7\times 10^4$ \\ [0.5ex] 
%
Red (13 Gyr, MF08), $V < 25.8$ 
                               &  low-$\rho_{\rm h}$ bin &   3 & $3.4\times 10^4$ & $3.3\times 10^4$ \\
                               & high-$\rho_{\rm h}$ bin & 115 & $1.5\times 10^5$ & $1.3\times 10^5$ \\
                               &      inner $\Rgal$ bin &  35 & $1.0\times 10^5$ & $9.1\times 10^4$ \\
                               &      outer $\Rgal$ bin &   7 & $4.9\times 10^4$ & $4.7\times 10^4$ \\ [1ex]
%
Red (13 Gyr, BM03), $V < 25.8$ 
                               &  low-$\rho_{\rm h}$ bin &   2 & $2.1\times 10^4$ & $2.0\times 10^4$ \\
                               & high-$\rho_{\rm h}$ bin & 125 & $1.3\times 10^5$ & $1.1\times 10^5$ \\
                               &      inner $\Rgal$ bin &  26 & $9.0\times 10^4$ & $0.8\times 10^5$ \\
                               &      outer $\Rgal$ bin &   5 & $4.3\times 10^4$ & $4.1\times 10^4$ \\ [0.5ex] 
\tableline
\multicolumn{3}{c}{~} \\ [-1.5ex]              
\end{tabular}
\tablecomments{Column (1): Star cluster sample and limiting $V$ magnitude. 
`MF08' and `BM03' refer to models used to apply 10 Gyr of dynamical evolution to 
the cluster sample in question. Only surviving clusters are considered. 
See discussion in \S\S\ \ref{s:massfunc} and \ref{s:evol_MFs}. 
(2): Star cluster subsample. 
(3): Median half-mass density in M$_{\odot}$ pc$^{-3}$. 
(4): Median of best-fit mass lost per cluster in M$_{\odot}$. 
The typical formal uncertainty of $\widetilde{\Delta}$ returned by the fits 
is $\sim$\,25\%.  
(5) Peak mass associated with turnover of luminosity function, in M$_{\odot}$. 
}   
\end{center}
\end{table*}

\clearpage



\end{document}